\documentclass[reprint,NumberedRefs]{JASA}

\usepackage{amsmath,amssymb,amsfonts}
\usepackage{graphicx}
\usepackage{textcomp}
\usepackage{xcolor}
\usepackage{xfrac}
\usepackage{enumitem}
\usepackage{bm}
\usepackage{url}
\usepackage{notation}
\usepackage[normalem]{ulem}
\usepackage{mathtools}
\usepackage[ruled,vlined]{algorithm2e}

\newcommand{\stkout}[1]{\ifmmode\text{\sout{\ensuremath{#1}}}\else\sout{#1}\fi}

\newcommand{\ist}{\hspace*{.3mm}}
\newcommand{\rmv}{\hspace*{-.3mm}}

\newcommand{\spacel}{\ist l}

\newcommand{\symB}{\ist\text{b}}
\newcommand{\symT}{\ist\text{t}}
\newcommand{\symBT}{\ist\text{bt}}
\newcommand{\symU}{\ist\text{u}}

\newcommand{\varB}{\sigma^{\symB}}
\newcommand{\varU}{\sigma^{\symU}}

\newcommand{\F}{\mathcal{F}}
\newcommand{\FB}{\mathcal{F}^{\symB}}
\newcommand{\FBT}{\mathcal{F}^{\symBT}}
\newcommand{\SetStriation}{\Set{L}}
\newcommand{\gTransformed}{\underline{g}}
\newcommand{\SB}{\rv{s}^{\symB}}
\newcommand{\ST}{s^{\symT}}
\newcommand{\PB}{\rv{p}^{\symB}}
\newcommand{\PT}{p^{\symT}}
\newcommand{\YBT}{\rv{y}^{\symBT}}
\newcommand{\ybtr}{y^{\symBT}}

\newcommand{\BScalar}{\eta}
\newcommand{\PScalar}{\alpha}
\newcommand{\GScalar}{\gamma}

\newcommand{\rML}{\hat{r}^\text{ML}}
\newcommand{\rf}{r\text{-}f}

\DeclareMathOperator*{\argmax}{arg\,max}

\begin{document}

\title{Statistical Model and Estimation Method for Ranging a Moving Ship Using a Single Acoustic Receiver in Shallow Water\\
}

\author{Junsu Jang}
\email{jujang@ucsd.edu}
\author{William S. Hodgkiss}
\author{Florian Meyer}
\email{flmeyer@ucsd.edu}
\affiliation{Marine Physical Laboratory, Scripps Institution of Oceanography, University of California San Diego, La Jolla, CA, 92093, USA}


\begin{abstract}
Passive acoustics is a versatile tool for maritime situational awareness, enabling applications such as source detection and localization, marine mammal tracking, and geoacoustic inversion. This study focuses on estimating the range between an acoustic receiver and a transiting ship in an acoustically range-independent shallow water environment. Here, acoustic propagation can be modeled by a set of modes that are determined by the shallow water waveguide and seabed characteristics. These modes are dispersive, with phase and group velocities varying with frequency, and their interference produces striation patterns that depend on range and frequency in single-hydrophone spectrograms. These striation patterns can often be characterized by the waveguide invariant (WI), a single parameter describing the waveguide’s properties. This paper presents a statistical model and corresponding WI-based range estimation approach using a single hydrophone, leveraging broadband and tonal sounds from a transiting ship. Using data from the Seabed Characterization Experiment 2017 (SBCEX17), the method was evaluated on two commercial ships under different environmental conditions and frequency bands. Range estimation errors remained below $\pm 4$\% up to 62~km in the best case, with robust performance demonstrated in the 40–60 Hz band.
\end{abstract}

\maketitle

\section{Introduction}
Range information of a moving object is critical for maritime situational awareness. Passive acoustic ranging is an important task in maritime surveillance \cite{DreTraSiMicTes:J23,ZoBHilFra:J24} or in unmanned underwater vehicle (UUV) operations\cite{KitRanDiBSch:J22}. In addition, passive acoustic ranging can be used for self-localization by small autonomous platforms with limited navigation capabilities but equipped with a single hydrophone or a small hydrophone array. Such platforms include drifters\cite{JafFraRobMirSchKasBoc:J17}, gliders\cite{GraVanWorDziHow:J22,GehBarJohShaNolDav:J23} and propeller-driven UUVs\cite{PalLinFisKukKnoWil:C19}. While these systems rely on acoustic signals for localization, estimating the range of a non-cooperative, broadband acoustic source, such as a transiting commercial ship, using a single hydrophone in shallow water remains a significant challenge due to multipath propagation and modal interference.

Among single-hydrophone-based ranging methods, various approaches face distinct limitations. Matched field processing\cite{JesPorSteDemRodCoe:J00,CheKir:J22} relies either on detailed \textit{a priori} environmental knowledge on seabed properties and sound-speed profiles or the availability of signal replicas. Modal filtering-based methods\cite{BonTho:J14,LeGSocBon:J19} are restricted to impulsive or frequency-modulated signals, limiting their applicability to broadband ship noise. Cepstrum-based methods \cite{TraBarDreBou:J23,WatStiTesMey:J25} estimate time differences of arrival by extracting propagation path pairs. However, they are typically effective only for short-range estimations, on the order of a few kilometers. Deep learning approaches\cite{NiuGonOzaGerWanLi:J19} offer data-driven solutions but struggle with generalization and require extensive training data. In contrast, waveguide-invariant (WI)-based methods estimate range by processing interference patterns (striations) in spectrograms, enabling long-range passive tracking without detailed environmental knowledge. The WI, therefore, provides a powerful framework for passive acoustic ranging using a single receiver\cite{CocSch:J10,RakKup:J12,TaoHicKroKem:J07,TurOrrRou:J10,HarOdoKro:C15b,YouHarHicwRogKro:J20,SunGaoZhaGuoSonLi:J23,JanMey:C23,JanMey:C24}.


This paper extends WI-based ranging by introducing a statistical framework that jointly models broadband and tonal ship noise, with the potential of enhancing robustness against intensity variations and environmental uncertainty. 

\begin{figure}
    \centering
    \includegraphics[width=\reprintcolumnwidth]{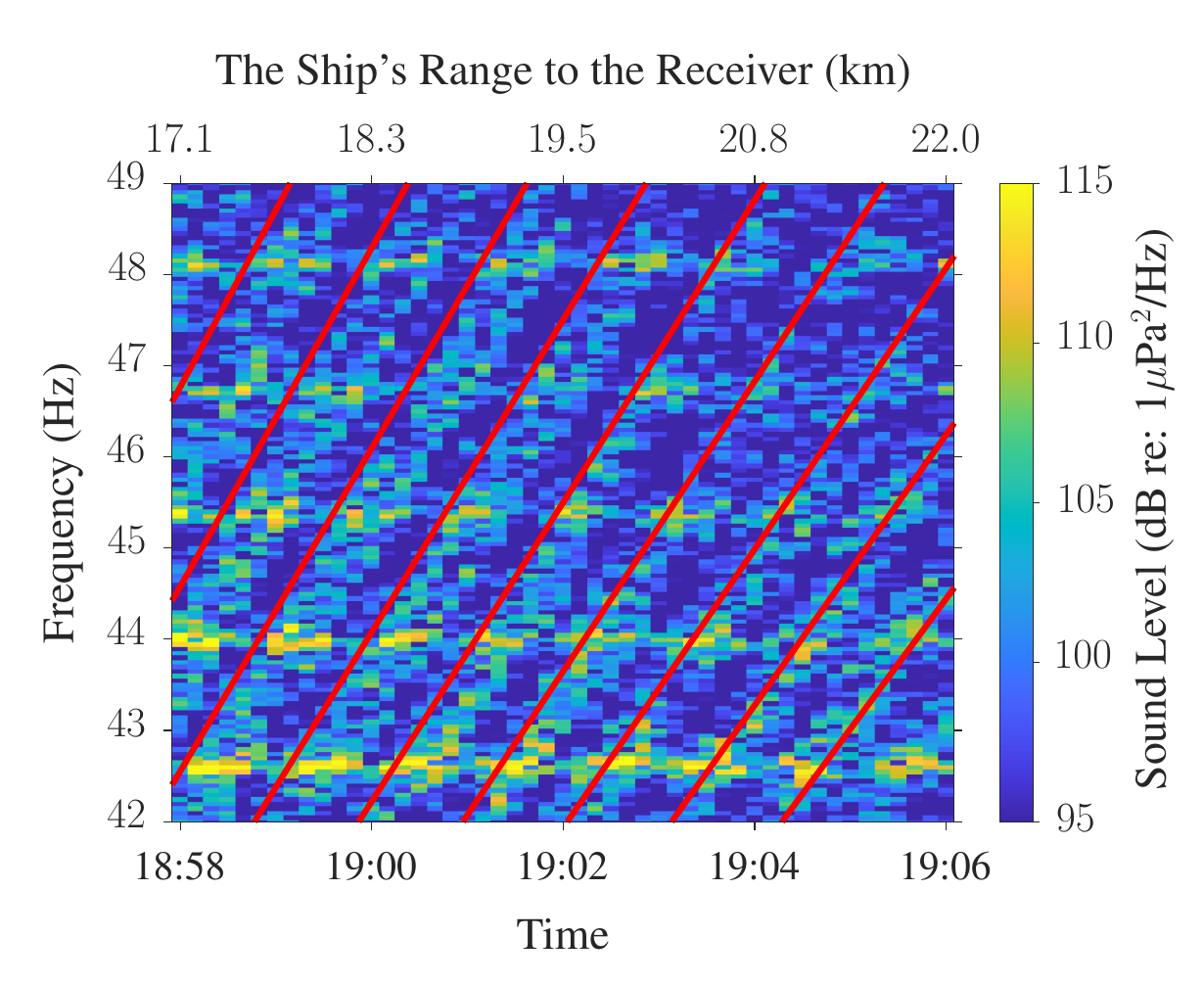}
    \caption{Spectrogram of data recorded on March 24, 2017 (UTC) during SBCEX17 using a single element of a vertical line array. A large commercial ship, MSC KALAMATA, was moving away from the receiver. Slightly diffuse tonal lines and broadband radiated noise from the ship are visible, along with modal interference patterns. The red lines show projected striations based on Eq.~\eqref{eq:wiStriation} using a waveguide invariant $\beta = 1.18$. Striations refer to interference-induced curves along which the acoustic intensity may follow local minima, maxima, or intermediate levels.}
   \label{fig:shipSoundEx}
\end{figure}

\subsection{Waveguide Invariant}\label{sec:wi}
In shallow water waveguides, sound propagates as discrete modes\cite{JenKupPorSch:B11_WI}. When an acoustic source moves relative to a receiver, interference between these modes produces striation patterns in a range-frequency representation of the received acoustic intensity. In an idealized setting, e.g., uniform source spectrum and negligible background noise, these striations manifest as interference-induced curves along which the acoustic intensity may follow local maxima, minima, or intermediate levels. These patterns, shaped by the waveguide environment, carry information about the acoustic source’s range, which can be extracted by making use of the WI\cite{Chu:J82,DspKup:J99,CocSch:J10b,CocSch:J11,Har:J11}. 

The WI, often denoted as $\beta$, is a parameter that describes how frequency-dependent modal interference relates to the source-receiver range. It arises from the functional relationship between the phase and group speeds of a group of modes\cite{ByuSon:J22}. It is typically defined for frequency bands sufficiently above mode cutoff frequencies, since dispersion properties change rapidly near the cutoff. While $\beta$ is typically treated as a single, mode-independent (scalar) value, in certain cases, e.g., when signals with low frequencies are used and few modes propagate, or when groups of modes with distinct dispersion characteristics are present, it may be better represented as a distribution derived from pairwise modal interactions\cite{RouSpi:J02,LeGBon:J13b}. For a comprehensive discussion on WI theory and practical considerations, see Refs.~\citen{JenKupPorSch:B11_WI} and~\citen{ByuSon:J22}.

Since the striations are functions of range and frequency, their shape can be used to infer the source’s range relative to the receiver. The WI establishes a mathtematical relationship between frequency and range, governing the shape of the striations. This relationship is given\vspace{-1mm} by
\begin{equation}\label{eq:wiStriation}
    \frac{f}{f'} = \biggl(\frac{r}{r'}\biggr)^{\beta}\rmv\rmv, 
\end{equation} 
where ($r', f'$) and ($r, f$) are different points along a striation, and $\beta$ is the WI. Since striations follow this relationship, a known WI allows for the projection of candidate striations for different candidate source ranges directly onto a spectrogram. This, however, requires mapping the time axis of the spectrogram to range using an assumed range rate first. The comparison of projected striations with observed striations in the spectrogram provides a direct method for inferring the source range. 


An example dataset for range estimation is shown in Fig.~\ref{fig:shipSoundEx}. It is a spectrogram of real acoustic data recorded during Seabed Characterization Experiment 2017 (SBCEX17)\cite{WilKnoNei:J20}, where a large commercial ship was moving away from the acoustic receiver in a $75$ m deep water column (This data will be discussed in more detail in Sec.~\ref{sec:data}). The tonal (horizontal lines of high intensities) and broadband components of ship-radiated noise are visible, along with striations formed by modal interference patterns. The red lines represent projected striations generated using Eq.~\eqref{eq:wiStriation} with the correct source ranges reported by the Automated Identification System (AIS) and the established WI value of $\beta = 1.18$. The red lines indicate striations, i.e., interference-induced curves in the spectrogram along which acoustic intensity remains approximately constant across frequency and time (range), with local maxima and minima most visually apparent. If incorrect ranges were used for projecting the line, the lines would no longer follow the intensity pattern in the spectrogram.

This paper introduces a statistical approach that quantifies how well the observed intensities along projected striations match modeled intensities provided by our statistical models of the received ship-radiated noise. Since the curvature of projected striations depends on the assumed source range, our estimation method selects the candidate range whose projected striations yield the best statistical agreement with the observed data.


\subsection{Ship Sound Characteristics}
The underwater noise radiated by transiting ships contributes significantly to the ocean ambient sound at low frequencies\cite{HavAdaHatVanDziHaxHepMcKMelGed:J21,Wen:J62}. It has been shown that, in the vicinity of the ship, high-intensity noise from 10 Hz to 1 kHz is emitted by the normal operation of the ship\cite{McKRosWigHil:J12}. Considering the relatively low attenuation at lower frequencies\cite{Uri:B13Ch5} and acoustic propagation in shallow water\cite{JenKup:J83}, low-frequency ship noise can be detected at ranges of tens of kilometers \cite{ScrHei:J91}.


The sound emitted from modern cargo ships can be considered as the superposition of the tonal and broadband acoustic signals\cite{Uri:B13Ch5} (see Fig.~\ref{fig:shipSoundEx}). The narrowband tonals correspond to the line spectrum that result from the periodicity of the machinery, such as the engine and the shaft\cite{ZhuGagMakRat:J22}. The broadband source signal stems from the propeller cavitation noise and flow noise around the hull, which are complex hydrodynamic processes and challenging to model accurately\cite{Ros:B76,WalHei:J02}. Depending on the physical properties of the ships, the mode of operations, the oceanographic conditions, and the aspect ratio, the sound level characteristics can vary\cite{McKWigHil:J13}. 

\subsection{Contribution and Paper Organization}
This paper introduces a new statistical model and method for estimating the range of a moving ship in shallow water using its radiated broadband and tonal acoustic signature, recorded by a single stationary hydrophone. By leveraging the WI, ranging is performed with minimal environmental knowledge through the analysis of the spectrogram of the received signal. A key assumption, as in prior work, is that the environment is range-independent, ensuring that the WI remains constant over range. Additionally, the WI of the considered environment and the ship's range rate, which is needed to convert the spectrogram's time-axis into range for each candidate range, are assumed to be known.

The proposed method addresses key limitations of existing WI-based ranging techniques. Deterministic methods\cite{CocSch:J10,RakKup:J12,YaoSunLiuJia:J21} that rely on directly extracting the slopes of the striations are sensitive to noise and require striations that exhibit relatively uniform intensity. The proposed statistical framework accounts for measurement uncertainties and variability in intensity along striations, improving robustness. Additionally, existing statistical methods focus on utilizing either the broadband\cite{JanMey:C24} or the tonal\cite{YouHarHicwRogKro:J20} components of the ship noise. By incorporating both components, our method increases the amount of usable data, reducing estimation uncertainty and improving range accuracy. The major contributions of the presented research are as follows.\vspace{-1mm}
\begin{itemize}
    \item \textbf{Statistical Models:} We develop a joint model for broadband and tonal ship sound in a range-independent shallow water environment.\vspace{-1.5mm}
    \item \textbf{Estimation Method:} We introduce a statistical signal processing approach that utilizes the WI for range estimation from spectrogram data.\vspace{-1.5mm}
    \item \textbf{Experimental Validation:} We evaluate the method on two commercial ships using real data from SBCEX17, demonstrating robust performance across frequency bands and channels, with range estimation errors under $\pm4$\% up to 62~km.
\end{itemize}


The paper is organized as follows: The general strategy for range estimation using the WI is introduced in Sec.\ref{sec:methodology}. The proposed statistical signal model for broadband and tonal noise components is established in Sec.~\ref{sec:SigModel}. Sec.~\ref{sec:rEst} describes the proposed method for range estimation using the model above. Data and the range estimation setup details are outlined in Sec.~\ref{sec:data} and Sec.~\ref{sec:estSetup}, respectively. Finally, results are discussed in Sec.~\ref{sec:results}, and Sec.~\ref{sec:conclusion} concludes the paper.

\section{Waveguide Invariant-based Ranging}\label{sec:methodology}

\begin{figure*}[t]
   \raggedleft
   \begin{minipage}{0.325\textwidth}
      \centering
      \centerline{\includegraphics[width=\linewidth]{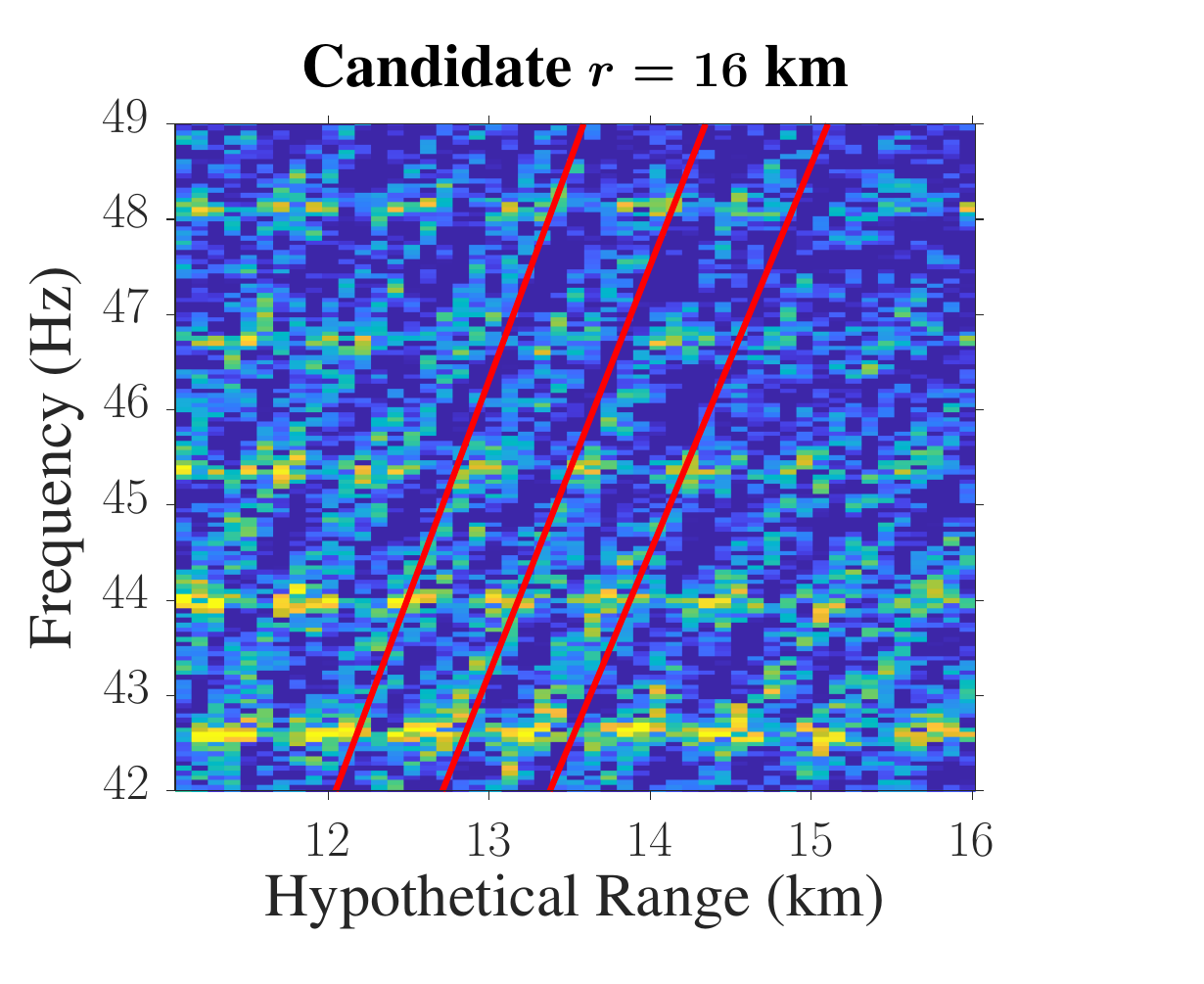}}
      \vspace{-2mm}
      \centerline{\small (a)}  
      \vspace{-2mm}
    \end{minipage}
   \begin{minipage}{0.325\textwidth}
      \centering
      \centerline{\includegraphics[width=\linewidth]{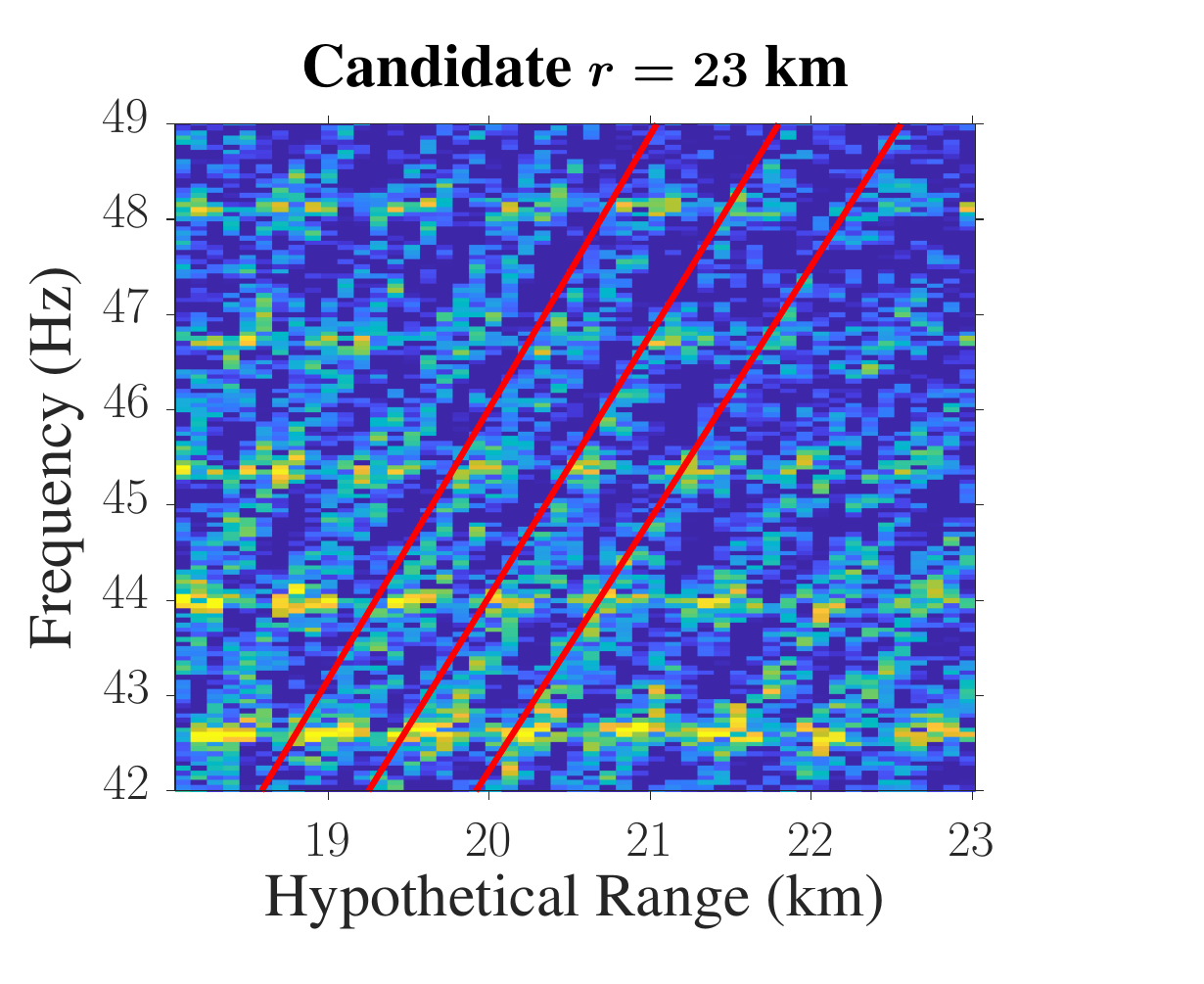}}
      \vspace{-2mm}
      \centerline{\small (b)}
      \vspace{-2mm}
   \end{minipage}
   \begin{minipage}{0.325\textwidth}
      \raggedright
      \centerline{\includegraphics[width=\linewidth]{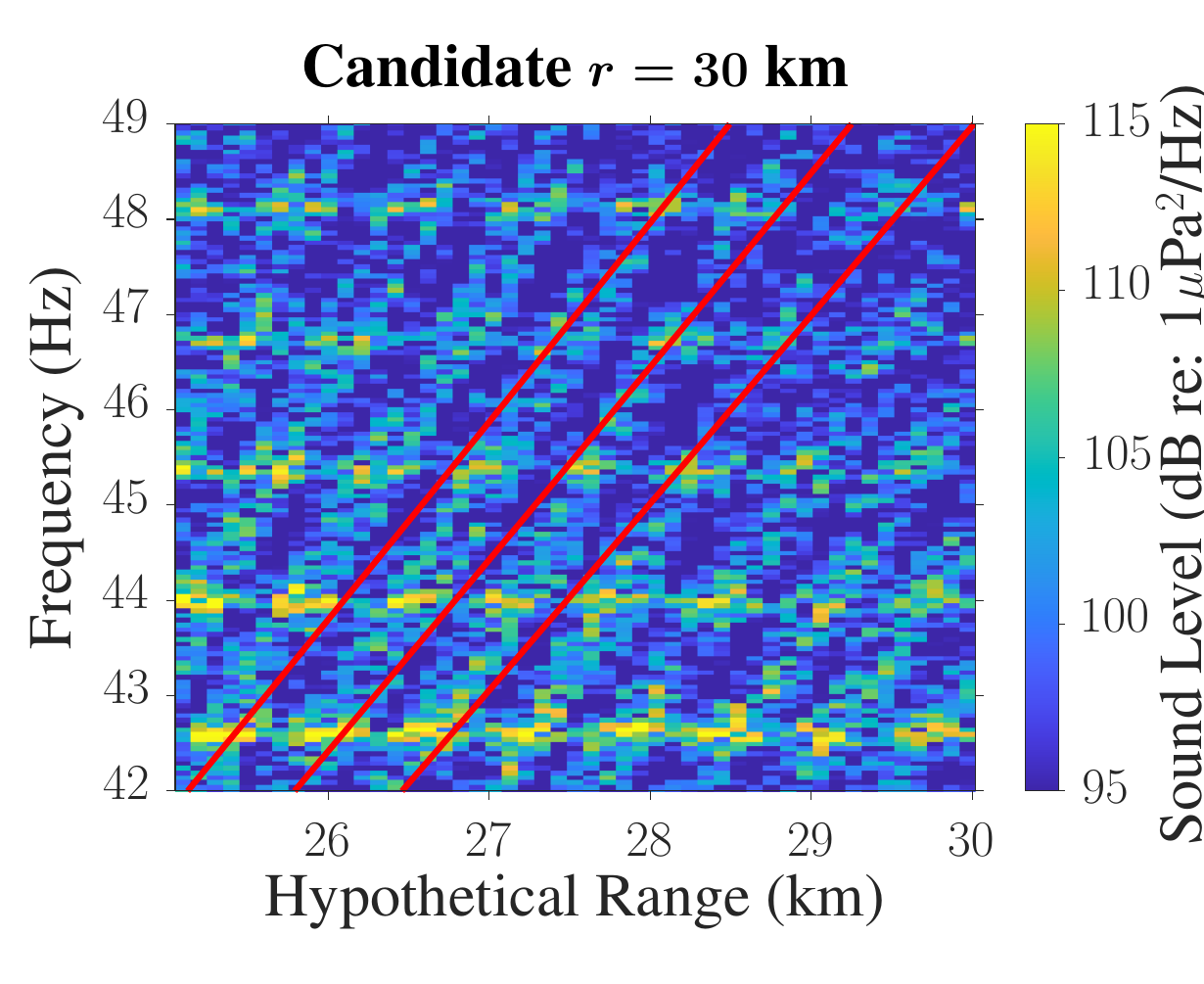}}
      \vspace{-2mm}
      \centerline{\small (c)}
      \vspace{-2mm}
   \end{minipage}
   \caption{Examples of the $\rf$ surface plots with projected striations (red lines) overlaid using a single reference frequency $f'=45.5$ Hz and Eq.~\eqref{eq:wiStriation}, for three candidate source ranges, which correspond to the final snapshot of the spectrogram. The candidate values are (a) $r=16$ km, (b) $r=23$ km, and (c) $r=30$ km. All plots are based on the same spectrogram shown in Fig.~\ref{fig:shipSoundEx}. For each case, only three striations are shown for clarity; in practice, a denser set of projected striations that span the full frequency band are included (see Sec.~\ref{sec:sfDomain}). The constant range rate $\dot{r}=10.2$ m/s and the WI $\beta=1.18$ are used. The true range is $23$ km, and the projected striations are most well aligned for (b), where the candidate range matches the true value.}
   \label{fig:striationProjections}
\end{figure*}

Our general strategy for estimating the source range is to systematically evaluate how well the expected striation patterns based on a set of candidate source ranges match those observed in a spectrogram. As in previous work on WI-based ranging\cite{TaoHicKroKem:J07,TurOrrRou:J10,HarOdoKro:C15b,YouHarHicwRogKro:J20,SunGaoZhaGuoSonLi:J23,JanMey:C23,JanMey:C24}, the source range estimation is based on Eq.~\eqref{eq:wiStriation}. Similarly to Ref.~\citen{YouHarHicwRogKro:J20}, we develop a statistical model and quantify the match between candidate striations and spectrogram based on a likelihood function. However, the statistical model used in Ref.~\citen{YouHarHicwRogKro:J20} is limited by the fact that only narrowband tonal frequencies are assumed to contain energy radiated by the ship, and neighboring frequency bins are modeled as range-independent background noise. The broadband component of the ship noise is not explicitly modeled and is effectively treated as a part of this background noise. Depending on the ship's operational and environmental context, such an approximation may be reasonable. However, the aforementioned method was validated using synthetically generated tonal signals that were constructed to conform to the model's assumptions, particularly the absence of broadband energy.

In our recording of ship noise, we observe that the broadband component carries non-negligible energy and also exhibits interference patterns (see Fig.~\ref{fig:shipSoundEx})\cite{VerSarCorKup:J17}. If this broadband content is treated as stationary background noise, it can lead to model mismatch and reduced ranging performance. More importantly, by disregarding the broadband component in ship noise, potentially useful information is neglected. Our proposed method incorporates both the tonal and broadband components into a unified statistical model, estimating their contributions directly from data and jointly leveraging them for source ranging. Furthermore, in the absence of ship-generated broadband energy, our model naturally reduces to the case considered by Ref.~\citen{YouHarHicwRogKro:J20}, relying solely on tonal content and background noise. This ensures compatibility with the original framework while extending its applicability of WI-based ranging to a broader range of real-world scenarios. Both our method and that of Ref.~\citen{YouHarHicwRogKro:J20} are applied to the same ship noise data to assess their respective performance and robustness under realistic conditions.

\subsection{Role and Estimation of WI and the Range Rate}\label{sec:methodologyWIRR}
To perform range estimation, knowledge of range rate $\dot{r}$ and WI $\beta$ is required due to the coupling of these parameters in Eq.~\eqref{eq:wiStriation} ($\dot{r}$ is needed to map time to range used in Eq.~\eqref{eq:wiStriation}). Since striation patterns are smooth functions of range and frequency, multiple parameter combinations can produce similar-looking striation patterns, leading to ambiguity in estimation. Approaches for obtaining knowledge on the WI $\beta$ and the range rate $\dot{r}$ to perform range estimation are reviewed in what follows.

\textit{\textbf{WI}}: In shallow water, the WI $\beta$ can often assumed to be $\beta = 1$ when the dominating group of modes are surface- and bottom-reflected\cite{ByuSon:J22}. Alternatively, the value of $\beta$ can be obtained in a calibration step by using ships of opportunity that share their position and velocity vector via the AIS \cite{VerSarCorKup:J17}.

\textit{\textbf{Range Rate}}: For this study, we used either a constant range rate based on AIS-reported ship speed or a time-varying rate computed from AIS position data.  However, a potential approach for estimating source range rate is to process acoustic recordings near the closest point of approach (CPA)\cite{TaoHicKroKem:J07,TurOrrRou:J10,SunGaoZhaGuoSonLi:J23}, if stable tonal components can be identified. Since the Doppler shift is zero at the CPA, the nominal tone frequencies can be extracted and used to estimate range rate based on the observed Doppler shifts before and after the CPA\cite{Bur:B02}. Our dataset includes CPA recordings, and preliminary inspection revealed a few tonal components between 300–800~Hz that are persistent and exhibited frequency shifts consistent with expected Doppler shifts. Although Doppler shift-based range rate estimation by exploiting CPA was not considered in the present work, this preliminary inspection suggests that it is a viable direction for future analysis.

\subsection{Method Overview}
The input data is the spectrogram of the recorded acoustic signal, and we want to estimate the source range, which corresponds to the final snapshot of the spectrogram, corresponding to the most last windowed segment in the Short-Time Fourier Transform. The method is based on the following key assumptions:
\begin{itemize}
    \item The WI $\beta$ and the range rate $\dot{r}$ are either estimated or approximated and therefore treated as known.
    \item The environment is range-independent, ensuring that $\beta$ remains constant over the observed range.
\end{itemize}
Errors in $\beta$ or $\dot{r}$ can introduce bias in the range estimate, as different parameter combinations may yield similar striation patterns, leading to ambiguity. 

Computing a range estimate for fixed $\beta$ and $\dot{r}$ is discussed next. For each range in a set of candidate source ranges, $r\in\Set{S}_r\rmv=\rmv [r_{\text{min}},r_{\text{max}}]$, the following steps are performed.
\begin{enumerate}
\item \textbf{Spectrogram Processing}: Using $r$ and $\dot{r}$, the time axis of the spectrogram is converted into a range axis, producing a hypothetical range-frequency ($\rf$) surface plot.

\item \textbf{Striation Projection}: For that hypothetical $\rf$ surface plot, the expected intensities along WI-projected striations using $\beta$ and Eq.~\eqref{eq:wiStriation} are interpolated on a regular range-frequency grid. 

\item \textbf{Likelihood Function Computation}: The joint likelihood function, parameterized by $\V{q}=[r,\dot{r},\beta]^{\text{T}}$, is evaluated. The likelihood function quantifies how well the observed intensity measurements in the striation-frequency domain match the projected striations based on  $\V{q}$ within the proposed statistical model.
\end{enumerate}
The most probable range is found by maximizing the joint likelihood function over all candidate ranges. 

Fig.~\ref{fig:striationProjections} shows examples of projected striation patterns overlaid on the same spectrogram for three different candidate source ranges. These $r$–$f$ surface plots illustrate how the alignment between projected striations and the observed intensity patterns varies with the assumed source range. Among the three candidates, the projection at $r = 23$ km exhibits the best alignment (Fig.~\ref{fig:striationProjections} (b)), consistent with the true source range. While the general striation patterns are visible in the spectrograms, due to the presence of the high intensity tonal signals and randomness in the broadband component as well as the background noise, the intensities are not uniform along each striation. These components are statistically modeled to enable evaluation of match between the projected and observed striation through the likelihood-framework.

The next section (Sec.~\ref{sec:SigModel}) establishes the statistical model for the measured intensities, laying the foundation for the proposed likelihood function in Sec.~\ref{sec:rEst}.

\section{Signal Model}\label{sec:SigModel}
This section introduces the statistical model of the received acoustic signal while a ship passes by an acoustic receiver in a range-independent shallow water environment. The tonal components of the ship radiated sound are treated as continuous sinusoidal signals with well-defined frequencies, whereas the broadband component is modeled as a random process due to its stochastic nature. Only a scenario where a single ship passes by the receiver is considered, and it is assumed that ambient noise sources are relatively distant and incoherent. 

The model is provided for the range-frequency domain representation of the acoustic signal, i.e., the $\rf$ surface plot. The $n^{\text{th}}$ snapshot corresponds to the acoustic signal at range $r_n$ for $n\in \mathcal{N} = \{1,\dots,N\}$, where $N$ is the number of snapshots in the $\rf$ surface plot. A discrete set of $K$ frequency bins, $\F = \{f_1,\dots,f_K\}$ with $f_1 < f_2 < \cdots < f_K$, is processed, where the $k^{\text{th}}$ frequency bin is $f_k \in \F$ for $k\in\Set{K} = \{1,\dots,K\}$.

The radiated acoustic signature of a moving ship is modeled as a superposition of broadband and tonal signals. The frequency bins are thus divided into two subsets: $\FB\subseteq\F$, containing only broadband components, and $\FBT\subset\F$, containing both broadband and tonal components. The broadband signal spans the entire band $\F$, while $\FBT$ consists of $J$ frequencies of narrowband tonal signals. For future reference, we define the corresponding index sets: $\Set{K}^{\text{b}} = \big\{k \ist \big| \ist f_k \in  \FB \big\}$ and $\Set{K}^{\text{bt}} = \big\{k \ist \big| \ist f_k \in  \FBT \big\}$ with $\Set{K} = \Set{K}^{\text{b}} \cup \Set{K}^{\text{bt}}$.

\subsection{Notation}
In what follows, the following notations are used. Random variables are displayed in sans serif, upright fonts, while their realizations are in serif, italic fonts. 
Vectors and matrices are denoted by bold lowercase and uppercase letters, respectively. For example, a random variable and its realization are denoted by $\rv x$ and $x$; a random vector and its realization by $\RV x$ and $\V x$; and a random matrix and its realization are denoted by $\RM{X}$ and $\M{X}$. 
The probability density function (PDF) and the expectation of random vector $\RV{x}$ are denoted as 
$f(\V x)$ and $\mathbb{E}[\hspace{.6mm}\RV{x}\hspace{.6mm}]$, respectively. Furthermore, $\mathcal{N}(u;\mu,\sigma^2)$ denotes a real Gaussian PDF (of random variable $\rv{u}$) with mean $\mu$ and variance $\sigma^2$, $\mathcal{CN}(x;\mu_{\text{c}},\sigma_{\text{c}}^2)$ denotes a complex Gaussian PDF  (of random variable $\rv{x}$) with mean $\mu_{\text{c}} \in \mathbb{C}$ and variance $\sigma_{\text{c}}^2\in \mathbb{R}$ such that each real and imaginary part has variance $\sigma_{\text{c}}^2/2$, $\Set{\chi}^2 (y; d, \lambda )$  denotes the noncentral chi-squared PDF (of random variable $\rv{y}$) with degrees of freedom $d$ and noncentrality parameter $\lambda$,  $\text{Exp}(z; \theta)$ denotes the exponential PDF  (of random variable $\rv{z}$) with scale parameter $\theta$.

\subsection{Received Signal Model}\label{sec:recSigModel}
At range $r_n$ and frequency $f_k \in \F$, the broadband and tonal source signals radiating from the moving ship are denoted as $\SB_{n,k} \in\mathbb{C}$ and $\ST_{n,k} \in\mathbb{C}$, respectively. The broadband source signal $\SB_{n,k}$ is modeled as $\mathcal{CN}\big(s^{\mathrm{b}}_{n,k}; 0,(\varB_k)^2\big)$, with frequency-dependent\cite{ZhuGagMakRat:J22} variance $(\varB_k)^2$. Nonrandom tonal source signals radiating from the moving ship are modeled by the complex amplitude $\ST_{n,k} = m^{\text{t}}_{k} \hspace{.3mm} \text{e}^{j\phi_{n,k}}\rmv$, where $m^{\text{t}}_{k} = | \ST_{n,k}| $ is an unknown, range-independent source magnitude that varies with frequency and $\phi_{n,k}$ is an unknown phase term, following Ref.~\citen{YouHarHicwRogKro:J20}. 

Numerous studies have proposed models for the spectral structure of ship-radiated broadband noise, both for specific ship types and operational conditions, e.g., Ref. \citen{ArvVen:J00}, and across large ensembles of vessels\cite{WalHei:J02}. In this work, the broadband component of the ship noise is modeled as a random process with frequency-dependent variance and no deterministic phase relationship between frequency components or across successive snapshots. This assumption allows us to statistically capture the spectral energy distribution by estimating the total received power (broadband signal plus noise) at each frequency, thereby learning the spectral envelope directly from the data. Note that our model does not capture the directionality of ship-radiated noise, which is often nonuniform and varies across ship types, nor does it account for variations in the ship’s operational state (e.g., engine or shaft condition)\cite{ArvVen:J00,TreVasVag:J08}. The frequency-dependent complex Gaussian model may not fully represent such spatial and temporal dependencies, which can appear to be caused by range-dependent effects. However, it offers analytical tractability and effectively captures the spectral diversity observed in real ship signatures. Most importantly, as will be demonstrated later in this paper, it makes it possible to perform ranging of ships over long distances.

The received acoustic signal is modeled as 
\begin{equation}
\rv{z}_{n,k} = g_{n,k}(\SB_{n,k} + \ST_{n,k}) + \rv{u}_{n,k}. \label{eq:acousticME}
\end{equation} 
For a range-frequency index pair ($n,k$), $g_{n,k}\in\mathbb{C}$ is the Green's function, i.e., the channel transfer function, evaluated at ($n,k$), and $\rv{u}_{n,k}\in\mathbb{C}$ is a random additive background noise following the PDF $\mathcal{CN}\big(u_{n,k}; 0,(\varU_k)^2\big)$. The Green’s function governs the observed interference pattern, and the key assumption used in WI-based ranging is that the magnitude of the Green's function only varies slowly along the striations and that the variation is independent of the considered striation\cite{SonByu:J20,KimKimByuKimSon:J24}. The mathematical details of the Green's function are further provided in App.~\ref{app:channel}. The background noise is assumed to be independent and identically distributed (iid) across range index $n$, i.e., across spectrogram time bins, and independent but not identically distributed across frequency index $k$,  with variance $(\varU_k)^2$ dependent on frequency\cite{YouHarHicwRogKro:J20}.

The model in Eq.~\eqref{eq:acousticME} is simplified and distinguished for two subsets of frequencies $\FB$ and $\FBT$ as
\begin{equation}\label{eq:generalReceivedSignal}
\rv{z}_{n,k} = 
\begin{cases}
    \PB_{n,k} +\rv{u}_{n,k}                 & k \in \Set{K}^{\text{b}}.\\[1mm]
    \PB_{n,k} + \PT_{n,k} + \rv{u}_{n,k}    & k \in \Set{K}^{\text{bt}} \\[.5mm]
\end{cases}
\end{equation}
where $\PB_{n,k} = g_{n,k}\SB_{n,k}$ and $\PT_{n,k}=g_{n,k}\ST_{n,k}$. The received broadband signal, $\PB_{n,k}$, is random and follows the PDF $\mathcal{CN}\big(p^{\ist\symB}_{n,k}; 0,(\varB_k)^2|g_{n,k}|^2\big)$, while the received tonal signal, $\PT_{n,k}  \hspace{-.3mm} =  \hspace{-.3mm} g_{n,k} \hspace{.3mm} \ST_{n,k}$, is nonrandom and unknown. Due to range-dependency of $g_{n,k}$, the statistics of $\PB_{n,k}$ are also range-dependent.

Across all frequency bins, $k \in \Set{K}$, the broadband‐plus‐noise signal $\PB_{n,k} +\rv{u}_{n,k}$ is distributed according to $\mathcal{CN}\big(\rv{z}_{n,k};0,\sigma^2_{n,k}\big)$ with variance $\sigma_{n,k}^2$. This variance $\sigma_{n,k}^2$ will be referred to as the broadband-plus-noise variance, which can be expressed as\vspace{-.7mm}
\begin{equation}
    \sigma_{n,k}^2 = (\varB_k)^2|g_{n,k}|^2 + (\varU_k)^2. \label{eq:generalizedVariance}\vspace{-.7mm}
\end{equation}
In addition, it is assumed that $\varU_k$ and $\varB_k$ vary smoothly with frequency index $k$. This implies that $\varU_k$ is approximately equal to $(\varU_{k+1} + \varU_{k-1})/2$ and $\varB_k$ is approximately equal to  $(\varB_{k+1} + \varB_{k-1})/2$.

\subsection{Statistical Model of Acoustic Intensity}
To perform range estimation, we are interested in the acoustic intensity defined as\vspace{-.7mm}
\begin{equation}\label{eq:intensityMeas}
    \rv{x}_{n,k} = |\rv{z}_{n,k}|^2.\vspace{-.7mm}
\end{equation}
Consequently, for frequency bins with only the broadband component ($k\in\Set{K}^{\text{b}}$), the intensity $\rv{x}_{n,k}$ follows an exponential distribution with scale parameter $\theta_{n,k} = \sigma_{n,k}^2$, i.e., $\rv{x}^{\text{b}}_{n,k} \sim \text{Exp}\big(x^{\text{b}}_{n,k} ; \theta_{n,k} \big)$. The intensity $\rv{x}^{\text{b}}_{n,k}$ is statistically independent across $n$ and $k$, and its scale parameter $\theta_{n,k}$ is both range- and frequency-dependent. Based on the functional form of $\text{Exp}\big(x^{\text{b}}_{n,k} ; \theta_{n,k} \big)$, the PDF of $\rv{x}^{\text{b}}_{n,k}$ is\vspace{-4mm}
\begin{equation}\label{eq:expPDF}
f_{\text{Exp}}(x^{\text{b}}_{n,k} ; \theta_{n,k}) = \frac{1}{\theta_{n,k}}\exp{\biggl(-\frac{x^{\text{b}}_{n,k}}{\theta_{n,k}} \biggr)}.
\end{equation}
For future reference, let $\M{X} \in (\mathbb{R}^{+})^{N\times K}$ be the measurement matrix, i.e., the $\rf$ surface plot, that consists of elements $x_{n,k}$, where $n\in\mathcal{N}$ and $k\in\Set{K}$.

For frequency bins with broadband and tonal component, $k \in \Set{K}^{\text{bt}}$, based on the assumptions above, the received acoustic signal follows a circularly symmetric complex Gaussian distribution, $\rv{z}_{n,k}=\PB_{n,k} + \PT_{n,k} + \rv{u}_{n,k} \sim \mathcal{CN}(z_{n,k}; \PT_{n,k},\sigma_{n,k}^2)$, with mean $\PT_{n,k}$ and variance $\sigma_{n,k}^2$ (cf. Eq.~\eqref{eq:generalizedVariance}). Note that both  $\PT_{n,k}$ and $\sigma_{n,k}^2$ are range- and frequency-dependent. Let us introduce the normalized intensity defined\vspace{-2mm} as
\begin{equation}\label{eq:ybt}
    \YBT_{n,k} = \frac{\rv{x}^{\symBT}_{n,k}}{\sigma_{n,k}^2/2},
\end{equation}
where $\rv{x}^{\symBT}_{n,k}$ is the intensity at frequency index $k\in\Set{K}^{\symBT}$. The normalized intensity follows a noncentral-$\chi^2$ PDF with two degrees of freedom\cite{EvaHasPea:B00Ch9}, i.e., $d=2$, and noncentrality parameter $\lambda_{n,k} = |\PT_{n,k}|^2/\sigma_{n,k}^2$, i.e., $\rv{y}^{\symBT}_{n,k}\sim\chi^2\big(\rv{y}^{\symBT}_{n,k};2,\ist\lambda_{n,k}\big)$. Note that $|\PT_{n,k}| = m^{\text{t}}_{k} \hspace{.15mm} |g_{n,k}|$. The normalized intensity is assumed statistically independent across $n$ and $k$. Based on the functional form of $\chi^2\big(\rv{y}^{\symBT}_{n,k};2,\ist\lambda_{n,k}\big)$, the PDF of $\YBT_{n,k}$ can be written as
\begin{equation}\label{eq:nc2PDF}
f_{\chi^2}(\ybtr_{n,k};\lambda_{n,k}) = \frac{1}{2}\exp{\bigg( \hspace{-1.5mm}-\frac{\ybtr_{n,k}+\lambda_{n,k}}{2} \bigg)} \hspace{.5mm} I_0\bigg(\sqrt{\ybtr_{n,k} \ist \lambda_{n,k}}\bigg),
\end{equation}
where $I_0(\cdot)$ is a modified Bessel function of the first kind and zeroth order. 

To perform range estimation with the given signal model, however, the values of the introduced statistical parameters must be known. It is assumed that the statistics of the ambient noise are available from the acoustic measurements when the ship was not present, i.e., the background noise variance $(\varU_k)^2$ is known \textit{apriori} at the time of processing. To compute the normalized intensity measurements $\YBT_{n,k}$, we need to obtain the broadband-pluse-noise variance $\sigma_{n,k}^2$, which will be  computed from the measurement matrix $\M{X}$. To evaluate the PDFs of $\YBT_{n,k}$, we also need to know the noncentrality parameter $\lambda_{n,k}$, which will also be computed from the measurement matrix $\M{X}$. The computations of $\lambda_{n,k}$ and $\sigma_{n,k}^2$ are provided in App.~\ref{app:estDetails}.

Based on numerical studies (not shown), the estimators for $\lambda_{n,k}$ and $\sigma_{n,k}^2$ can yield small biases, particularly when only a small number of tonal lines (e.g., two or three) are used. In the simulations, the Green's function is computed using a normal mode propagation model with a specified sound speed profile and geoacoustic parameters representative of the environment. For range estimation, however, we adopt the simplified model in App.~\ref{app:channel}, in which the Green’s function magnitude varies only with frequency along each striation. This simplification does not capture the simulated Green’s function magnitude dependence on range, and the resulting mismatch leads to the observed biases. However, despite these biases, range estimation remains accurate in practice. This is because the striation patterns projected under incorrect range hypotheses lead to a significantly larger mismatch between modeled and observed intensity structure and, in turn, to a significantly lower likelihood value. The correct range produces striations that better align with the underlying modal interference, yielding a significantly larger likelihood value.

\subsection{Signal Model in Striation-Frequency Domain}\label{sec:sfDomain}
To facilitate statistically principled range estimation, we transform the received data from the range-frequency domain to the striation-frequency domain, where acoustic intensities are sampled along projected striation curves defined by Eq.~\eqref{eq:wiStriation}. The striation-frequency domain serves as a physics-informed coordinate system in which the statistics of intensity and normalized intensity measurements along each striation are analytically tractable for inference.

The transformation is carried out using a candidate parameter vector $\bm{q} = [r, \dot{r}, \beta]^\text{T}$. In the transformed domain, the original range axis of the spectrogram is replaced by a striation index $l \in \SetStriation = \{1, \dots, L\}$, which corresponds to projected striations. Specifically, for a fixed reference frequency $f'\in \F$ and each reference range $r'$ along the range axis, a striation curve is generated by computing range $r$ across frequencies $f_k \in \F$ that satisfy Eq.~\eqref{eq:wiStriation}. Only those striations that span the full frequency band of interest are retained, and the striation axis in the transformed domain is defined by these valid projections. (See Fig.~\ref{fig:striationProjections} for a visual illustration of the projected striations, which define the striation axis in the transformed domain. In our implementation, the frequency in the middle of the band is chosen as the reference frequency.)

We define the intensity vector for a given striation as $\V{\underline{x}}_{\spacel}(\bm{q}) = \big[ \underline{x}_{\spacel,1 \hspace{-.4mm}}(\bm{q}), \dots, \underline{x}_{\spacel,K \hspace{-.4mm}}(\bm{q}) \big]^{\text{T}}$, where $\underline{x}_{\spacel,k}(\V{q})$ is obtained by interpolating measurements from the $\rf$ surface plot along the $l^\text{th}$ striation. We assume that after this transformation, intensities and normalized intensities remain statistically independent across $k$ and $l$, i.e., signal models in Sec.~\ref{sec:SigModel} also hold for $n$ replaced by $l$. For future reference, a variable with an underline notation, i.e., $\underline{x}$, indicates that the variable, i.e., $x$, has been transformed using Eq.~\eqref{eq:wiStriation}. 

\section{The Proposed Range Estimation Method}\label{sec:rEst}
We aim to estimate the source range $r$ by maximizing a joint likelihood function that, for a candidate parameter vector $\bm{q}$, describes the match between modeled intensities and measured intensities statistically. This framework also allows for the estimation of WI $\beta$ when the range and range rate are known, or range rate $\dot{r}$ when the range and WI are known. We therefore define a general likelihood function in terms of a joint parameter vector $\V{q} = [r,\dot{r},\beta]^{\text{T}}$. This formulation allows us to solve for range or WI or constant range rate within the same estimation framework. 

We evaluate the PDFs of both the intensity measurements $\underline{x}^{\symB}_{\spacel,k}(\V{q})$ at frequencies $f_k\in\FB$ with only the broadband components and the normalized intensity measurements $\underline{y}^{\symBT}_{\spacel,k}(\V{q})$ at frequencies $f_k \in \FBT$ with both broadband and tonal components across all striations $l\in\SetStriation$. The joint likelihood function of $\V{q}$ given the intensity measurements  $\underline{x}^{\symB}_{\spacel,k}(\V{q})$ at frequencies $f_k\in\FB$ is defined as
\begin{equation}\label{eq:lB}
\ell^{\symB}( \V{q} ) = \prod_{k\in\Set{K}^{\symB}} \hspace{.25mm} \prod_{l\in\SetStriation} \hspace{.5mm} f_{\text{Exp}}\big(\hspace{.25mm} \underline{x}^{\symB}_{\spacel,k}(\V{q});\hat{\underline{\theta}}_{\spacel,k} (\V{q})\hspace{.25mm} \big),
\end{equation}
and the joint likelihood function of $\V{q}$ given the normalized intensity measurements $\underline{y}^{\symBT}_{\spacel,k}(\V{q})$ at frequencies $f_k\in\Set{F}^{\symBT}$ is defined as
\begin{equation}\label{eq:lBT}
\ell^{\symBT}( \V{q} ) = \prod_{k\in\Set{K}^{\symBT}} \hspace{.25mm} \prod_{l\in\SetStriation} \hspace{.5mm} f_{\chi^2}\big(\hspace{.25mm} \underline{y}^{\symBT}_{\spacel,k}(\V{q});\hat{\underline{\lambda}}_{\spacel,k} (\V{q})\hspace{.25mm} \big).
\end{equation}
The scaling factor $\hat{\underline{\theta}}_{\spacel,k}$ and the noncentrality parameter $\hat{\underline{\lambda}}_{\spacel,k}$ are estimated from the measurements. The functional form of $f_{\text{Exp}}(\cdot)$ and $f_{\chi^2}(\cdot)$ are provided in Eqs.~\eqref{eq:expPDF} and \eqref{eq:nc2PDF}, respectively. Finally, the joint likelihood function of $\V{q}$ given both intensity and normalized intensity measurements in their respective frequencies is
\begin{equation}
\ell( \V{q} ) = \ell^{\symB}( \V{q} )\ist\ist\ell^{\symBT}( \V{q} ).
\end{equation}

We aim to find the parameter vector $\V{q}$ that best aligns the measured intensities and normalized intensities with their modeled distributions. To estimate $\V{q}$, we seek the most probable set of parameters that maximize the likelihood. This follows a standard ML estimation framework, where the optimal estimate is given\vspace{-.5mm} by
\begin{equation}\label{eq:JLL_general}
\hat{\bm{q}}^{\text{ML}} = \argmax_{\bm{q} \in \mathcal{S}} \hspace{1mm}\ell( \V{q} ),
\end{equation}
where $\Set{S}$ denote the set of candidate parameter vectors $q$ over which the likelihood is evaluated. In practice, the logarithm of the likelihood, i.e., the log-likelihood, is maximized for numerical stability. 

As discussed in Sec.~\ref{sec:methodologyWIRR}, unambiguous estimation of $\bm{q} = [r,\dot{r},\beta]^{\text{T}}$ is only possible if the uncertainty for two of the three parameters is small. To estimate range based on Eq.~\eqref{eq:JLL_general}, it is assumed that  the range rate $\dot{r}$ and the WI $\beta$ in the joint parameter vector $\V{q}$ are known. In practice, we estimate them using separate methods (see Sec.~\ref{sec:methodologyWIRR}) and use their estimates $\hat{\beta}$ and $\hat{\dot{r}}$ in the likelihood function. When performing range estimation, this leads to the practical formulation $\V{q}=[r,\hat{\dot{r}},\hat{\beta}]^{\text{T}}$ which is optimized over the candidate source ranges $\Set{S}_r$. The estimate is given by
\begin{equation}
\hat{r}^{\text{ML}} = \argmax_{r \in \mathcal{S}_{r}} \hspace{1mm}\ell( \V{q}=[r,\hat{\dot{r}},\hat{\beta}]^{\text{T}}).
\end{equation}
Similarly, if the range and the range rate are treated as known (estimated as $\hat{r}$ and $\hat{\dot{r}}$), the WI is estimated by optimizing over candidate WI set $\Set{S}_\beta$ as
\begin{equation}\label{eq:bEst}
    \hat{\beta}^{\text{ML}} = \argmax_{\beta \in \mathcal{S}_{\beta}} \hspace{1mm}\ell(\V{q}=[\hat{r},\hat{\dot{r}},\beta]^{\text{T}}).
\end{equation}

Note that likelihood evaluation is performed in the striation-frequency domain corresponding to a candidate parameter vector $\V{q}$. The key challenge is obtaining the statistical parameters necessary to evaluate the likelihood function. We outline their computation in App.~\ref{app:estDetails}, and key steps are provided in Algorithm~\ref{algo:MLE_rA} in the same appendix.

\section{Experimental Environment and Data Collection}\label{sec:data}
\begin{figure}[t]
    \centering   
    \begin{minipage}{.9\reprintcolumnwidth}
        \centering
        \centerline{\includegraphics[width=\reprintcolumnwidth]{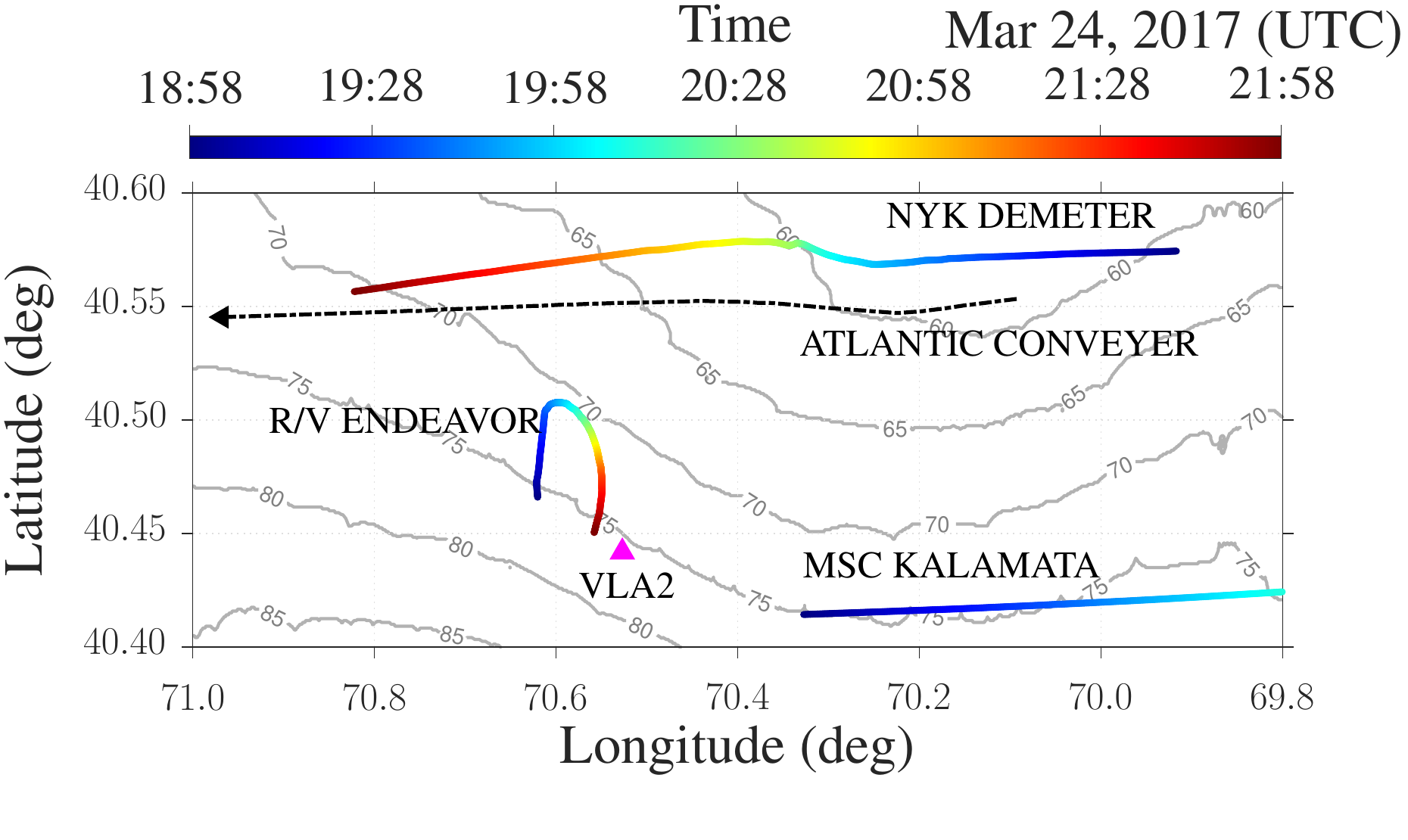}}
        \vspace{-3mm}
    \end{minipage}
    \caption{The considered ship tracks in the vicinity of vertical line array 2 (VLA2, triangle) during SBCEX17, overlaid on bathymetric contours (gray lines). The color-coded trajectories correspond to the time window from 18:58 to 21:58 UTC on March 24, 2017, showing the passage of MSC KALAMATA, NYK DEMETER, and R/V ENDEAVOR. An additional ship track of westbound ATLANTIC CONVEYER (a black dashed line) occurred earlier in the same day. Its sound recording was used to obtain the WI used for estimating the NYK DEMETER ranges.}
   \label{fig:SBCEX17event}
\end{figure}

The acoustic data used in this study were collected during the SBCEX17 experiment, conducted near the New England Mud Patch on the U.S. continental shelf. All analyzed recordings were made by the second vertical line array (VLA2), deployed by the \textit{Marine Physical Laboratory} at a nominal water depth of 76~m. The array comprised 16 hydrophones spanning the full water column. Several large commercial ships transited nearby shipping lanes during the experiment: an eastbound lane south of VLA2 and a westbound lane to the north.

The environment is considered relatively range-independent, with mild bathymetric variation (within 10~m over tens of kilometers) and a weakly stratified sound speed profile, with a nominal value of 1473~m/s, measured near VLA2. The seabed consists of fine-grained mud overlying layered sediments, but geoacoustic properties such as density and layer thickness are expected to vary spatially along the ship tracks~\cite{BalGarLeeVenMcNWilCha:J24}. 

\subsection{Ship Tracks and Acoustic Events}\label{sec:shipTracks}
\begin{figure}[t]
   \begin{minipage}{\reprintcolumnwidth}
      \centering
      \centerline{\includegraphics[width=\reprintcolumnwidth]{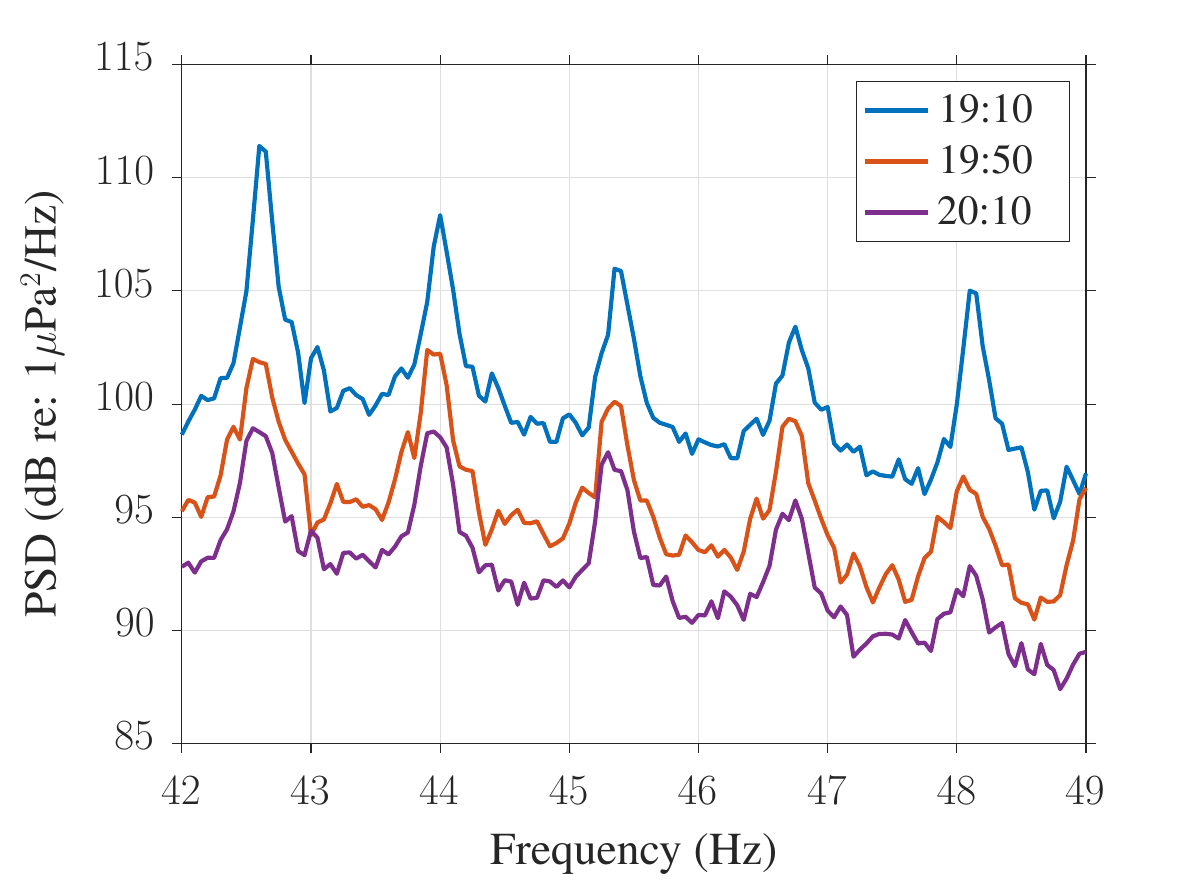}}
   \end{minipage}
    \caption{Progression of the power spectral density (PSD) in the 42–49~Hz band, computed from 10-min long spectrograms recorded by the 8\textsuperscript{th} element of VLA2 on March 24, 2017 (UTC). Each curve corresponds to a PSD during the MSC KALAMATA transit, with spacing of 30 minutes. Five strong tonal components associated with the ship's acoustic signature remain clearly visible throughout, even as the ship receded beyond 60~km.}
   \label{fig:SBCEX17_zoomIn}
\end{figure}

This study focuses on a subset of transits between 18:58 and 22:25 UTC on March 24, 2017, during which two vessels—MSC KALAMATA (MMSI: 477510600) and NYK DEMETER (MMSI: 353025000)—passed within acoustic detection range of VLA2. These ships were selected due to their strong tonal and broadband signatures. Fig.~\ref{fig:SBCEX17_zoomIn} shows the progression of the power spectral density (PSD) measured at the 8\textsuperscript{th} element (41.25~m depth) of VLA2 as MSC KALAMATA moved away from the array. Despite increasing range, five distinct tonal peaks remained visible throughout the event, highlighting the persistence of tonal energy critical to the range estimation methods evaluated in this study. 

Fig.~\ref{fig:SBCEX17event} summarizes the analyzed ship tracks, overlaid on the regional bathymetry. The locations of the ships and therefore their ranges relative to VLA2 used as ground truths throughout this study are processed from the AIS reports recorded locally on R/V ENDEAVOR (MMSI: 303471000) and Marine Cadastre\cite{cadastre:W24}. R/V ENDEAVOR also transited the area at low speed (average 3.4~kn) but contributed minimally to the acoustic interference below 60 Hz, which is of interest for range estimation. Thus, the interference of R/V ENDEAVOR was deemed minimal for this study. An additional vessel, ATLANTIC CONVEYER (MMSI: 266018000), transited the northern lane earlier in the day; its acoustic signature was used to estimate the WI required for estimating the range of NYK DEMETER (see Sec.~\ref{sec:impl_WIRR}). 

\begin{figure*}[t]
    \centering
    \centerline{\includegraphics[width=\textwidth]{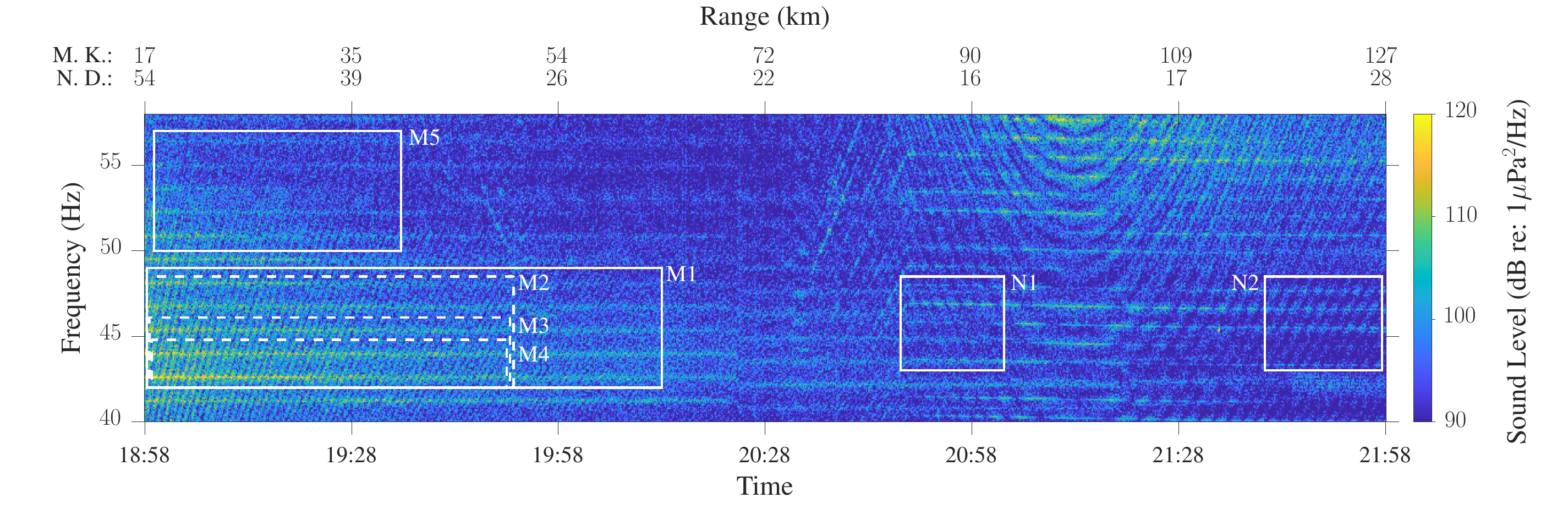}}
    \caption{Spectrogram of the processed acoustic recordings from the 8\textsuperscript{th} element of VLA2 over a time window on March 24, 2017 (UTC), showing the radiated acoustic signatures of MSC KALAMATA and NYK DEMETER. The corresponding source ranges for each ship are indicated at the top axis for reference. NYK DEMETER only began to transit at a steady speed at 20:48, after which its acoustic signature became suitable for ranging. The scenario is discussed further in Sec.~\ref{sec:shipTracks}. Panels M1–M5 denote the frequency bands used for range estimation of MSC KALAMATA, while N1–N2 indicate the spectrogram segments used for NYK DEMETER. See Sec.~\ref{sec:evalCond} for how the data corresponding to each panels are used.}
    \label{fig:spectrogramKD}
\end{figure*}

Fig.~\ref{fig:spectrogramKD} shows the spectrogram of the recordings from the 8\textsuperscript{th} element of VLA2 during the analysis window. MSC KALAMATA first appeared near VLA2 and transited eastward along the southern shipping lane at a steady speed of 19.8~kn. NYK DEMETER was traveling westward in the northern shipping lane and became closer to VLA2 at 19:31, reaching a range of 37~km. However, NYK DEMETER began to decelerate and change heading around 19:40, and its acoustic signature weakened. Clear tonal striations only reappeared after 20:48, when the vessel resumed steady westward motion at 16.1~kn. This change is visible in the spectrogram as a rising harmonic structure between 20:30 and 20:50, associated with vessel acceleration. 

\section{Range Estimation Setup}\label{sec:estSetup}
The statistical model assumes range-independent propagation and a single WI value over the duration analyzed. These conditions are met by design in the selected experimental segments for evaluation of the proposed method. The method’s empirical performance thus reflects both the suitability of the model and the underlying physical environment, as further discussed in Sec.~\ref{sec:results}. The implementation details on the spectrogram computation, frequency band selection, ambient noise and candidate range can be found in App.~\ref{app:impl}.

\subsection{Summary of Evaluation Conditions}\label{sec:evalCond}
Tab.~\ref{tab:evalSummary} summarizes the various conditions in which the proposed method is evaluated using the recordings of the two ship passages. To systematically evaluate the performance of the proposed WI-based range estimation method, we focus on the acoustic recordings of MSC KALAMATA, which provided clear tonal and broadband signatures over a wide range of distances. The main evaluation was performed using the spectrogram segment labeled M1 in Fig.~\ref{fig:spectrogramKD}, which spans a broad time window. This segment was divided into overlapping 2-minute intervals, each containing enough data to extract at least $L=30$ valid striation projections for all considered candidate ranges. These data were also processed using hydrophone recordings from each element of VLA2 to examine the impact of receiver depth.

\begin{table}[t]
\caption{Summary of range estimation evaluation conditions corresponding to the spectrogram panels in Fig.~\ref{fig:spectrogramKD} for the proposed method (G). The table lists the ship name (M.~K. for MSC KALAMATA, N.~D. for NYK DEMETER), spectrogram time window (UTC) on March 24, 2017, the corresponding panel identifier in Fig.~\ref{fig:spectrogramKD}, the waveguide invariant (WI, denoted $\beta$), processed frequency band, and receiver depth. For the first row, range estimation was performed using all 16 elements of VLA2 spanning from 15 to 70.75~m depth in 3.75~m intervals.}
\begin{adjustbox}{width=\reprintcolumnwidth}
\begin{tabular}{c|c|c|c|c|c}
\hline\hline
\begin{tabular}{@{}c@{}}Ship\vspace{-1.5mm}\\Name\end{tabular} & Time & \begin{tabular}{@{}c@{}}Panel\vspace{-1.5mm}\\ID\end{tabular} & WI, $\beta$ & \begin{tabular}{@{}c@{}}Bandwidth\vspace{-1.5mm}\\(Hz)\end{tabular} & \begin{tabular}{@{}c@{}}Receiver\vspace{-1.5mm}\\Depth (m)\end{tabular} \\
\hline
M. K. & 18:58-20:12 & M1 & 1.18 & 42-49 & 15-70.75 \\
M. K. & 18:58-19:46 & M3 & 1.18 & 42-45.6 & 41.25 \\
M. K. & 18:58-19:46 & M4 & 1.18 & 42-44.6 & 41.25 \\
M. K. & 18:58-19:46 & M5 & 1.18 & 50-57 & 41.25 \\
N. D. & 20:48-21:10 & N1 & 1.33 & 43-48.5 &  41.25 \\
N. D. & 21:35-21:58 & N2 & 1.06 & 43-48.5 & 41.25 \\
\hline\hline
\end{tabular}
\end{adjustbox}
\label{tab:evalSummary} 
\end{table}

The proposed method was benchmarked against two established approaches: the slope-based estimation method of Ref.~\citen{CocSch:J10} and the tonal-based statistical method from Ref.~\citen{YouHarHicwRogKro:J20}. All three methods were applied to the same acoustic data beginning at 18:58 UTC, and their performance was compared with respect to range estimation accuracy and the maximum distance at which each method remained reliable.

We also assessed the impact of frequency selection by processing two distinct frequency bands: 42–49~Hz and 50–57~Hz (panels M1 and M5 in Fig.~\ref{fig:spectrogramKD}, respectively). Both bands exhibited strong tonal signatures, but the ship acoustic signatures persisted longer in the lower frequency band. In addition, shorter spectrogram segments labeled M2–M4 were selected to assess range estimation using a reduced number of tonal components. The performance of the proposed method and the tonal-based statistical method\cite{YouHarHicwRogKro:J20} are compared in these conditions, highlighting the benefits of including broadband components in the proposed method.

To demonstrate the method’s generalizability across different environmental conditions, we also applied it to NYK DEMETER’s acoustic recordings by the 8\textsuperscript{th} hydrophone. Two fixed-length spectrogram segments labeled N1 and N2 in Fig.~\ref{fig:spectrogramKD} were selected to represent pre-CPA (20:48-21:10) and post-CPA (21:35-21:58) intervals, respectively. A slightly narrower frequency band (43-48.5~Hz) is processed due to strong tonal noise close to 42~Hz. These examples qualitatively show the method’s ability to perform range estimation under different propagation conditions, particularly in environments characterized by different WI values.

\subsection{WI and the Range Rate}\label{sec:impl_WIRR}
\textbf{\textit{WI}}: The WI value ($\beta=1.18$) used for range estimations of MSC KALAMATA is estimated from the earliest spectrogram (18:58-19:10 and 42-49 Hz band). For estimating the range of NYK DEMETER pre-CPA ($\beta=1.33$) and post-CPA ($\beta=1.06$), the WI values were derived from the recordings of ATLANTIC CONVEYER (41-50 Hz band) rather than NYK DEMETER. This choice was made because NYK DEMETER exhibited a non-straight trajectory with notable speed fluctuations pre-CPA, which introduced ambiguity in the interference patterns. In contrast, ATLANTIC CONVEYER maintained a more consistent speed and straight track during its transit through the northern region, making it more suitable for stable WI estimation (see Fig.~\ref{fig:SBCEX17_zoomIn}). 

The estimated values of $\beta$ deviate from the canonical shallow-water $\beta = 1$ expected under idealized iso-velocity conditions and flat bathymetry \cite{ByuSon:J22}. Our preliminary analysis suggests that within this frequency band and setting, the observed modal interference pattern is more appropriately characterized via individual mode-pair interactions rather than a single global $\beta$ value \cite{LeGBon:J13a}. In particular, only a single pair of modal pair interference is observed in the considered ranges.

\textbf{\textit{Range Rate}}: The source range rate $\dot{r}$ was assumed constant and equivalent to the AIS-reported speed for MSC KALAMATA, i.e., 19.8~kn or 10.2~m/s. For NYK DEMETER, however, the CPA distance was relatively large (approximately 12 km), and the range rate varied significantly even beyond 30 km from the array. Therefore, a time-varying $\dot{r}(t)$ series was derived from AIS-based ship positions. For each spectrogram segment (pre-CPA and post-CPA), only the portion of $\dot{r}(t)$ matching that segment’s time window was used in the estimation. It is important to note that while AIS data were used to inform $\dot{r}(t)$, the corresponding source range was not used in the estimation process. Range estimation was performed independently, treating the source location as unknown. A full treatment of jointly estimating both range and range rate will be considered in future work.

\subsection{Benchmark Methods Used for Comparison}\label{sec:comparison}
To assess the performance of the proposed statistical method, we compare it with two benchmark approaches used for WI-based passive ranging. The comparison is made by processing the acoustic recording of MSC KALAMATA only. Each group of methods is labeled by its acronym, and the proposed method is referred to as method (G). 

\textbf{\textit{Slope-based methods} (S)}: Assuming the presence of a broadband signal, a class of methods \cite{CocSch:J10,VerSarCorKup:J17,YaoSunLiuJia:J21} employs a two-dimensional discrete Fourier transform to analyze the range-frequency structure of acoustic striations, extracting the dominant slope as a key parameter for range estimation. These methods assume that striations are approximately linear within a short range-frequency window, enabling local slope approximation. The extracted slope, $\text{d}f/\text{d}r = \beta \ist\ist f / r$, directly relates to the source range through the differential form of the WI equation.

However, these methods are sensitive to intensity variations, especially in the presence of tonal components, and require uniform intensity along striations. Additionally, when $\beta \neq 1$, striation curvature introduces errors, necessitating careful selection of the window length to maintain accuracy \cite{CocSch:J10}. In our study, we implemented the method outlined in Ref.~\citen{YaoSunLiuJia:J21}.

\textbf{\textit{Tonal-based statistical method} (T)}: Another class of WI-based ranging methods statistically processes prominent ship noise features \cite{HarOdoKro:C15b,YouSolHic:C17b,YouHarHicwRogKro:J20,JanMey:C24}. Ref.~\citen{YouHarHicwRogKro:J20} proposed a statistical model and search-based estimation method for tonal signals radiated by a moving ship in shallow water. While the general estimation framework is equivalent to the proposed method (G), it assumes that the background noise dominates over the broadband component of the ship noise, making the intensity statistics in non-tonal frequency bands effectively range-independent. Thus, the background noise level used for intensity normalization and noncentrality parameter $\underline{\lambda}$ estimation are computed directly from the off-tonal broadband frequency bins within the spectrogram. 

Additionally, method (T) requires estimating the noncentrality parameters $\underline{\lambda}$ for each striation via a search over candidate values, which increases computational cost. In contrast, the proposed method (G) computes these parameters directly from the data using a closed-form expression (cf. Eq.~\eqref{eq:nc2Est}), resulting in substantial computational savings. In our implementation of (T), the noncentrality parameter was searched from 0 to 100 in steps of 2.5. Empirically, the search step size smaller than 2.5 did not provide significant accuracy gain. Lastly, the method (T) uses the same striation-frequency domain representation as method (G), since both methods evaluate acoustic intensities along projected striations within the same transformed spectrograms (Sec.~\ref{sec:sfDomain}). 

\section{Results}\label{sec:results}
Percentage error is used as a range-normalized metric. In WI-based ranging, uncertainty in measured intensities along striations introduces range errors that grow approximately linearly with distance, but percentage errors remain relatively constant, offering a consistent measure across ranges.

\subsection{MSC KALAMATA}\label{sec:kalamata}
\begin{figure}[t]
   \begin{minipage}{\reprintcolumnwidth}
      \centering
      \centerline{\includegraphics[width=\linewidth]{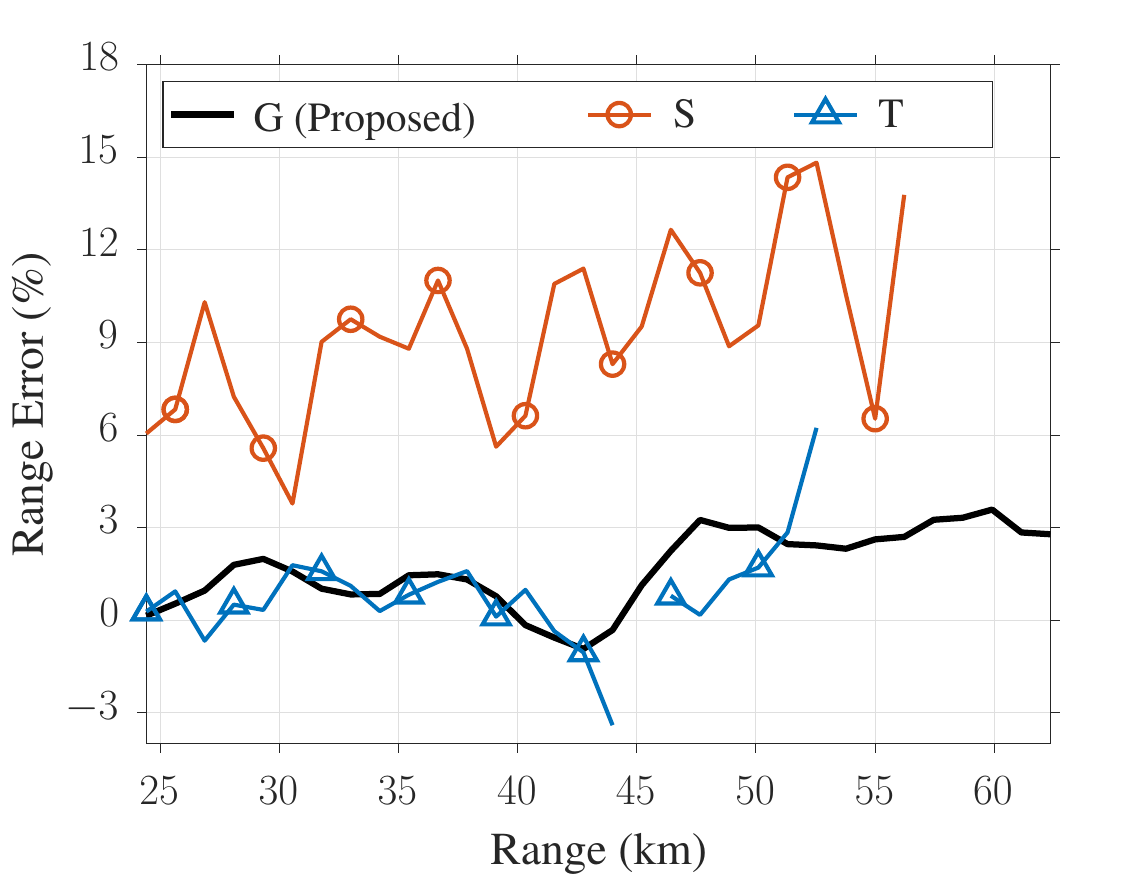}}
   \end{minipage}
   \caption{Range error (in percent) for three WI-based ranging methods: (G) proposed approach, (S) slope-based method, and (T) statistical approach using only tonal signals (see Sec.~\ref{sec:comparison} for details). Range estimates were computed every two minutes from the spectrogram segment labeled M1 (Fig.\ref{fig:spectrogramKD}) and smoothed using a centered moving average with a window size of 3. The plotted errors extend to the maximum range at which each method provided consistently reliable estimates. While method (S) exhibits a consistent positive bias, its error standard deviation remains stable up to 56~km. The anomalous estimation at 45~km for method (T) is excluded for visual clarity. The proposed method (G) maintained low error and variance across the entire range, performing reliably up to 62~km.}
\label{fig:rangeErr}
\end{figure}

\textbf{\textit{Method comparison}}: Fig.~\ref{fig:rangeErr} shows the range error as a function of distance for three methods: the proposed method (G), the slope-based method (S), and the tonal-only method (T). The analysis was conducted using segmented spectrograms from panel M1 in Fig.~\ref{fig:spectrogramKD}, covering the 42–49~Hz frequency band. The proposed method (G) consistently maintained sub-4\% error out to 62~km, with a root mean square error (RMSE) of 1.8\% up to 52~km. In comparison, the tonal-only method (T) exhibited a slightly higher RMSE of 2.5\% and was reliable only out to 52~km. The slope-based method (S) performed the worst, with an RMSE of 10.9\%. One anomalous estimate at 45~km was excluded for method (T).

\begin{figure}[t]
   \begin{minipage}{\reprintcolumnwidth}
      \centering
      \centerline{\includegraphics[width=\linewidth]{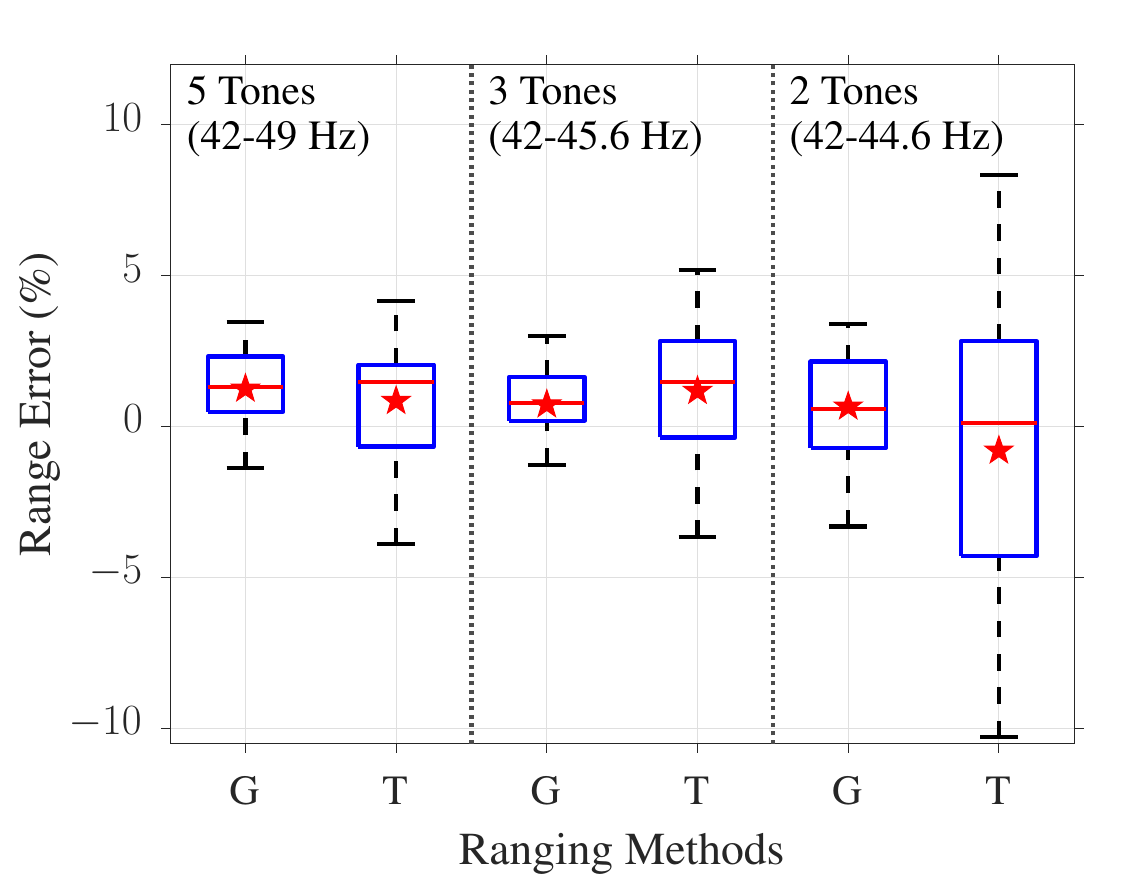}}
   \end{minipage}
   \caption{Box plot of range errors for MSC KALAMATA between 25~km and 52~km using the 42–49~Hz band. Results are shown for the proposed method (G) and the tonal-only method (T) across three frequency bands corresponding to different numbers of tonal components: five tones (M2), three tones (M3), and two tones (M4), as labeled in Fig.~\ref{fig:spectrogramKD}. In each box, the red line marks the median, the box edges indicate the 25th and 75th percentiles, and the red star denotes the mean. Whiskers extend to the most extreme data points, excluding one outlier in the 5- and 3-tone cases and two outliers in the 2-tone case for method (T). As the number of tones decreases, the performance of method (T) deteriorates significantly. In contrast, the proposed method (G) maintains low error and variability by leveraging both the broadband and tonal features of the ship signature.}
   \label{fig:boxPlot}
\end{figure}

The standard deviation of the error for (G) between 25–62~km was 1.4\%. The mean error increased from 0.9\% in the 25–45~km range to 2.9\% between 45–62~km. Since the ship maintained a steady heading and speed, the increasing bias suggests that the assumed WI $\beta = 1.18$ is slightly too large at long ranges, leading to a small positive bias in the estimated range.

The slope-based method (S) exhibited the largest bias and variability, with a mean error of 9.3\%, standard deviation of 5.7\%, and a maximum error of 19.4\%. Note that the peak value of the error is not fully shown in Fig.~\ref{fig:rangeErr} since we performed smoothing of the curves. The aforementioned statistics correspond to errors between 25 and 56 km. This rather poor performance stems from intensity variation across frequencies and the method’s sensitivity to striation curvature. In cases where $\beta \ne 1$, striations deviate from linearity, and fitting a straight slope to a curved pattern estimates a too small true slope, particularly when $\beta > 1$, leading to range estimates that are too large.

While the tonal-only method (T) capitalizes on high-SNR tonal components, it models broadband ship noise as range-independent background noise. This simplified model is used to normalize tonal intensity as well as estimate the noncentrality parameter. However, in our dataset, this simplified model is inaccurate: the broadband component of the ship noise also exhibits range-dependent structure due to modal interference, introducing strong striations outside the tonal bins. This mismatch degrades the performance of method (T), especially when broadband contributions are not negligible. The leftmost box plots in Fig.~\ref{fig:boxPlot} summarize the statistical performance of methods (G) and (T) across the 25–52~km range, confirming the advantages of incorporating broadband information.

To evaluate robustness under reduced tonal content, we also performed range estimation using subsets of the 42–49~Hz band: three tones (42–45.6~Hz) and two tones (42–44.6~Hz), within the 25–52~km range. We compared two methods: (G), which jointly models broadband and tonal components, and (T), which relies solely on tonal signals. The resulting error distributions are shown in Fig.~\ref{fig:boxPlot}. The proposed method (G) maintained consistent performance across all tonal subsets, while the tonal-only method (T) exhibited increasing variability and degraded accuracy as the number of tones decreased. These results highlight the benefit of incorporating broadband structure, especially when tonal features are weak or limited.

\begin{figure}[t]
   \begin{minipage}{\reprintcolumnwidth}
      \centering
      \centerline{\includegraphics[width=\linewidth]{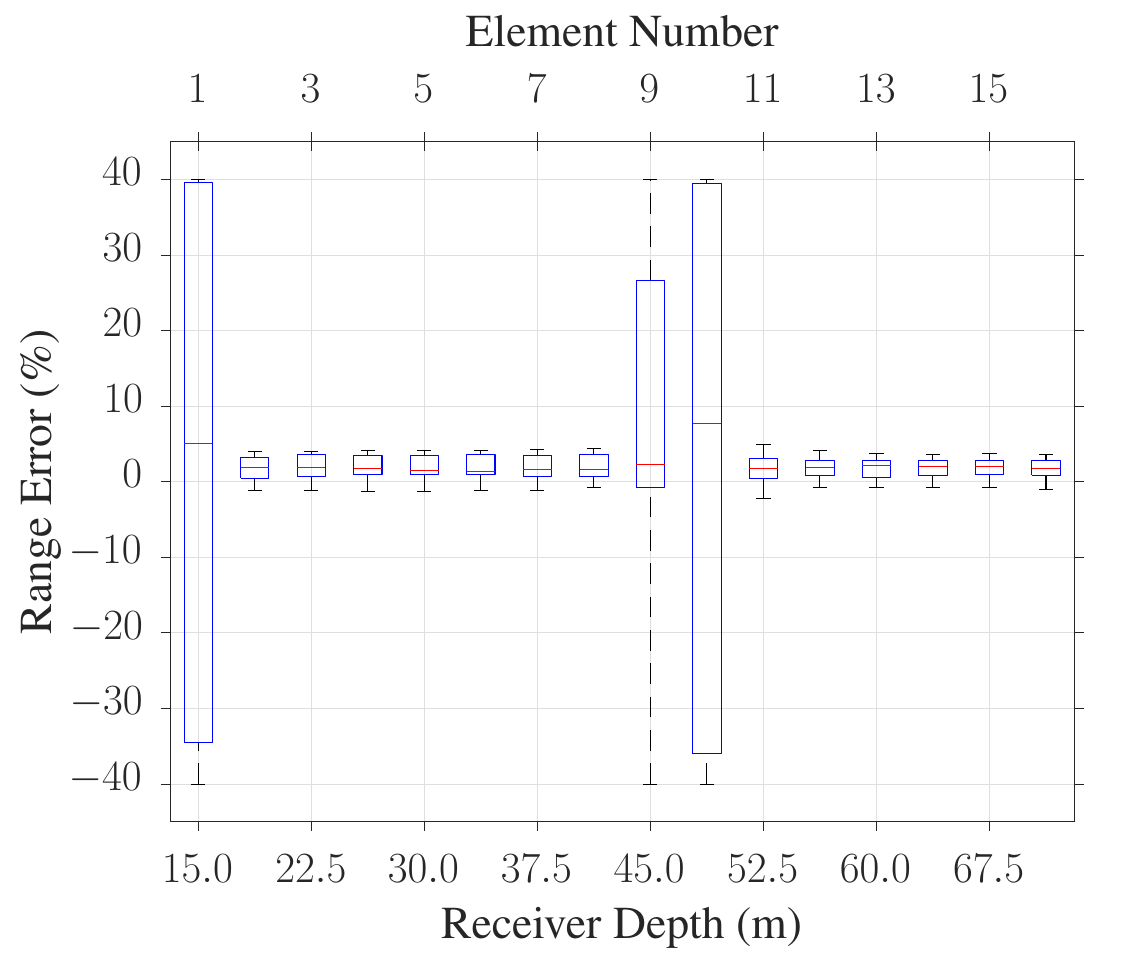}}
   \end{minipage}
   \caption{ Box plot of range errors for MSC KALAMATA between 25 and 62~km using the 42–49~Hz band and the proposed method (G), evaluated across all receivers of VLA2. Box plot conventions follow those described in Fig.~\ref{fig:boxPlot}. Five outliers for the 11\textsuperscript{th} element and one outlier each for the 12\textsuperscript{th} and 13\textsuperscript{th} elements were excluded from the plot. The hydrophone near the surface (1\textsuperscript{st} element) suffered from high ambient noise, while those near 50~m depth (9\textsuperscript{th}-11\textsuperscript{th}) coincided with mode nulls, especially the 10\textsuperscript{th} element, leading to degraded ranging performance.}
   \label{fig:receiverDepth}
\end{figure}

\textbf{\textit{Receiver Depth Sensitivity}}: Range estimation using the proposed method (G) was performed on recordings from all 16 hydrophones of VLA2, which span the full water column with 3.75~m spacing. Estimates were computed over the 42–49~Hz band for source ranges between 25 and 62~km. The resulting range errors are summarized as box plots in Fig.~\ref{fig:receiverDepth}. Most hydrophones yielded accurate and consistent results; however, performance degraded for the shallowest element (15~m depth) and for elements 9–11 (45–52.5~m depth). The 11\textsuperscript{th} element, in particular, produced five anomalous estimates excluded from the figure.

The shallow receiver was affected by persistent, high-intensity noise across the frequency band of interest, possibly due to surface dynamics or mechanical interference. At mid-depth (elements 9–11), although tonal energy remained strong, the striation patterns were weak or absent, limiting the performance of the method. This may be due to elevated ambient or instrument noise, but we hypothesize that these depths coincide with a vertical mode null in one of the interfering modes. Since striation patterns arise from interference between at least two propagating modes, a null in one mode can suppress the interference pattern altogether. Supporting this, interference patterns reappeared at higher frequencies ($>$80~Hz), where the modal structure is expected to differ. A more detailed analysis of these modal effects is beyond the scope of this paper.

\textbf{\textit{Frequency Band Comparison}}: We evaluated range estimation performance at the 8\textsuperscript{th} element using two 7-Hz-wide frequency bands: 42–49~Hz and 50–57~Hz, each containing five tonal components. In the higher-frequency band, accurate estimation was limited to 39~km, with a mean error of 2.8\% and a standard deviation of 3.6\% over the 25–39~km range. In contrast, the 42–49 Hz band yielded accurate estimates up to 62 km, with a mean error of 1.7~\% and a standard deviation of 1.4~\% over the 25–62 km range. The reduced performance at higher frequencies is attributed to lower SNR and weaker tonal visibility (see panels M1 and M5 in Fig.~\ref{fig:spectrogramKD}), which limited the ability to extract reliable striation patterns. These results underscore the importance of selecting frequency bands with strong radiated energy for robust and long-range estimation.

\subsection{NYK DEMETER}\label{sec:NDResults}
Range estimation for NYK DEMETER was performed using two spectrogram segments (one pre-CPA and one post-CPA) to illustrate the method’s performance under different propagation conditions characterized by distinct WI values. Each segment was analyzed independently, producing a single range estimate representative of its respective time window. In both cases, the spectrograms covered the 43–48.5 Hz band and contained four strong tonal lines. The WI values ($\beta=1.33$ pre-CPA and $\beta=1.06$ post-CPA) and the time-varying range-rate inputs for each segment were assigned as described in Sec.~\ref{sec:impl_WIRR}.

For the pre-CPA segment, a 22-minute spectrogram spanning 20:48–22:10 (panel N1 in Fig.~\ref{fig:spectrogramKD}) was processed using $\beta = 1.33$. The source range decreased from 18.1~km to 14.6~km in this time window.  The range estimate was 1.4~km (7.7\%) too small. This relatively large error is likely due to a mismatch between the WI value estimated from ATLANTIC CONVEYER (used as a reference) and the actual data along the NYK DEMETER track.

For the post-CPA segment, a 23-minute spectrogram from 21:35–21:58 (panel N2 in Fig.~\ref{fig:spectrogramKD}) was analyzed using $\beta = 1.06$. During this time, the vessel’s range increased from 19~km to 28~km. In this case, the range estimate was only 0.7~km (2.6\%) too large, suggesting better agreement between the assumed WI value and the actual propagation conditions.

\section{Conclusions and Future Work}\label{sec:conclusion}
This paper proposes a statistical signal processing method for estimating the range between a stationary single hydrophone and a moving ship in range-independent shallow water, leveraging the spatial-interference pattern on the spectrogram of the ship's radiated signature. This pattern is characterized by the WI. The proposed method was validated using real acoustic recordings from a vertical line array element deployed during SBCEX17, where large commercial ships transited along nearby shipping lanes. For the primary ship analyzed, the proposed approach achieved range estimation errors within $\pm4\%$ out to 62~km, exceeding the performance and range limits of both the slope-based and tonal-only methods. By integrating the strengths of statistical methods that process either broadband or tonal components, the proposed method achieved robust range estimation results.

The method was evaluated across two ship tracks, with acoustic recordings spanning different geographic regions, receiver depths, and frequency bands. These tests demonstrate the robustness of the proposed method under realistic conditions and support the feasibility of using WI values obtained from one ship to estimate the range of other ships transiting the same region. Across all test cases, region-specific WI values, estimated once using reference spectrograms, were successfully reused for subsequent range estimation, even in the presence of mild trajectory offsets and potential environmental variability.

While the results demonstrate promising performance, certain assumptions, and experimental constraints warrant further investigation. The statistical parameters used for likelihood evaluation, such as the broadband scaling factors, noncentrality parameters, and ambient background noise statistics, were estimated from the data and assumed to be perfect during inference. Errors related to these parameter estimates are currently not captured by our statistical model, which may affect the robustness of our approach. While the proposed method demonstrated high estimation accuracy on the dataset in different regions of the considered experimental site, further testing and a comparison with reference methods across diverse environmental conditions, would be valuable to assess general applicability in other shallow water environments.

Areas for further investigation include extending the approach to range-dependent environments, where bathymetric and/or sound speed variations may affect modal propagation and thus the striation structure. Incorporating range-dependent WI formulations would enhance the method's applicability across a broader range of ocean settings\cite{DspKup:J99,ByuSonCho:J21,KimByunSongKimSong:J25}. Although the method was applied successfully in the 40–60 Hz band, the complexity of modal interference at shorter ranges, particularly when multiple low-order modes are not strongly attenuated, may challenge the assumption of a single dominant WI. These effects could motivate more detailed modeling in future studies that use ship acoustic signatures in low-frequency bands. In contrast, at higher frequencies, the approximation $\beta = 1$ may be more valid\cite{ByuSon:J22}. Applying and validating the method in higher frequency bands could help assess its broader utility.

Finally, developing methods to estimate the range rate directly from the acoustic signal would remove the current need to obtain this information from an additional sensor. Although previous work has proposed range-rate estimation based on phase differences across ranges, e.g., Ref.~\citen{RakKup:J12}, our application of this method on ship-radiated noise did not yield reliable estimation results. Alternatively, a CPA-based range rate estimation approach that tracks multiple tones to extract Doppler shift information\cite{MeyKroWilLauHlaBraWin:J18,JanMeySnyWigBauHil:J23} could be explored in future studies.

\begin{acknowledgments}
We would like to thank Drs. Heechun Song, Donghyeon Kim and Gihoon Byun for helpful discussion on WI-based ranging. This work was supported by the Defense Advanced Research Projects Agency (DARPA) under Grant D22AP00151 and by the Office of Naval Research (ONR) under Grant N00014-23-1-2284. 
\end{acknowledgments}

\section*{Data Availability}\label{sec:DataAvailability}
The acoustic data that support the findings of this study are available from W. Hodgkiss upon reasonable request.

\appendix
\section{Estimation Details}\label{app:estDetails}
\subsection{The Green's Function}\label{app:channel}
Without loss of generality, let $(n,k)$ and $(n',k')$, with $n,n'\in\Set{N}$ and $k,k'\in\Set{K}$, denote the range and frequency index pairs that, according to Eq.~\eqref{eq:wiStriation}, lie on the same striation. A central assumption in WI-based ranging is that the magnitude of the Green's function changes slowly along each striation \cite{SonByu:J20,KimKimByuKimSon:J24}). Based on this assumption, we introduce the channel model\vspace{-1.5mm} 
\begin{equation}\label{eq:GScalarAssumption}
|g_{n,k}| = \GScalar_{k} |g_{n',k'}|.
\end{equation}
where the scaling factor $\GScalar_{k}\in \mathbb{R}^{+}$ is independent of the considered striation and only frequency-dependent. Furthermore, Eq.~\eqref{eq:GScalarAssumption} can be re-expressed in the transformed domain as
\begin{equation}
\label{eq:GScalarAssumptionTransformed}
|\gTransformed_{\spacel,k}| = \GScalar_{k} |\gTransformed_{\spacel,K}|,
\end{equation}
where $l$ is a striation index.

\subsection{Broadband-Plus-Noise Variance Estimation}\label{app:broadParam}
The broadband-plus-noise variance $\underline{\sigma}_{\spacel,k}^2$ at striation with index $l\in\SetStriation$ and a frequency with index $k\in\Set{K}^{\symB}$, defined in Eq.~\eqref{eq:generalizedVariance}, is required to compute the normalized intensity measurements which are then used within the likelihood function. This variance is estimated by determining the scale parameters of the intensity measurements using the relationship established in Sec.~\ref{sec:SigModel}: \begin{equation}\label{eq:scaleVarRelation} 
\underline{\theta}_{\spacel,k} = \underline{\sigma}_{\spacel,k}^2. 
\end{equation}

We leverage the fact that, under the transformation defined by the correct parameter vector $\V{q}$, the scale parameters of the exponentially distributed intensities within a striation (across frequency indices $k\in\Set{K}^{\symB}$) follow a structured relationship dictated by the Green's function (App.~\ref{app:channel}). Within our statistical model, at a reference frequency $f_{k'}\in\F$, the scale parameter varies across striations (cf. Eqs.~\eqref{eq:generalizedVariance} and \eqref{eq:scaleVarRelation}). At any other frequency $f_k\in\F\backslash\{f_{k'}\}$, the scale parameter for a given striation is a scaled version of the corresponding scale parameter at $f_{k'}$. The scaling factor is the same for all striations at a given frequency, as it depends only on the frequency-dependent variations in the magnitude of the source amplitude and the channel transfer function (cf. Eqs.~\eqref{eq:acousticME}, \eqref{eq:GScalarAssumption}, and  \eqref{eq:GScalarAssumptionTransformed}). These scaling factors can be estimated by comparing measured intensities across different frequencies.

In particular, computing the broadband-plus-noise variance $\underline{\sigma}_{\spacel,k}^2$ can be mathematically described as follows. For a given frequency index $k\in\Set{K}^{\symB}$, we define $\BScalar_{k} = \varB_k/\varB_{k'}$ ($\BScalar_{k} > 0$) as the ratio of source broadband signal standard deviations at $f_k$ and $f_{k'}$, where $k'\in\Set{K}^{\symB}$ is a fixed reference frequency index chosen arbitrarily. Consequently, the source signal can be modelled as $\rv{s}^{\symB}_{n,k}\sim\mathcal{CN}(0,(\varB_{k'})^2\BScalar_k^2\hspace{.3mm})$. Based on Eqs.~\eqref{eq:generalizedVariance} and \eqref{eq:scaleVarRelation},
\begin{equation}\label{eq:scaleParam}
    \underline{\theta}_{\spacel,k} = (\varB_{k}|\gTransformed_{\spacel,k}|)^2 + (\varU_{k})^2. 
\end{equation}
Using the definition of $\BScalar_{k}$ and Eq.~\eqref{eq:GScalarAssumptionTransformed}, it follows that
\begin{equation}\label{eq:GScalarEq2}
\underline{\theta}_{\spacel,k} = (\eta_k \gamma_k)^2 (\varB_{k'}|\gTransformed_{\spacel,k'}|)^2 + (\varU_{k})^2, 
\end{equation}
where $k'$ is an arbitrarily chosen reference frequency index. We introduce the received broadband scaling factor $\PScalar_{k} = (\eta_k \gamma_k)^2$ and the received broadband signal variance $\underline{v}_{\spacel,k} = (\varB_k)^2|g_{n,k}|^2$, for $k\in\Set{K}^{\symBT}$. Then, the scale parameter $\underline{\theta}_{\spacel,k}$ can be expressed as:
\begin{equation}
    \underline{\theta}_{\spacel,k} = \PScalar_{k}\underline{v}_{\spacel,k'} + (\varU_{k})^2.\label{eq:scaleParamExpanded}
\end{equation}

\textbf{\textit{Scaling Factor Estimation}}: To estimate $\underline{\theta}_{\spacel,k}$, we first compute the estimates $\hat{\alpha}_k$ of $\PScalar_{k}$ and $\hat{\underline{v}}_{\spacel,k'}$ of $\underline{v}_{\spacel,k'}$. The received broadband scaling factor $\PScalar_{k}$ is estimated as
\begin{equation}\label{eq:alphaB}
\hat{\alpha}_k = \frac{\frac{1}{M}\sum_{l\in\SetStriation}\underline{x}^{\symB}_{\spacel,k}- (\varU_{k})^2}{\frac{1}{M}\sum_{i\in\SetStriation}\underline{x}^{\symB}_{\hspace{.6mm} i,k'}-(\varU_{k'})^2}. 
\end{equation}
The a detailed derivation and justification of this estimator is provided in supplementary material\footnote{\label{fn:first}See
Supplementary materials at [URL will be inserted by AIP]
for the theoretical justification and derivation of the scaling factor and noncentrality parameter estimators.}. Additionally, the received broadband signal variance $\underline{v}_{\spacel,k'}$ is estimated as
\begin{equation}\label{eq:bbVarEst}
\hat{\underline{v}}_{\spacel,k'} = \frac{1}{\frac{1}{K-J}\sum_{k\in\Set{K}^{\symB}} \hat{\alpha}_k} \biggl(\frac{1}{K-J}\sum_{k\in\Set{K}^{\symB}}\underline{x}^{\symB}_{\spacel,k}- (\varU_{k})^2 \biggr).
\end{equation}
This estimator is also derived in detail in the supplementary material\footnotemark[\value{footnote}]. Finally, based on Eq.~\eqref{eq:scaleParamExpanded}, Eq.~\eqref{eq:alphaB} and Eq.~\eqref{eq:bbVarEst}, an estimate of $\underline{\theta}_{\spacel,k}$ is computed as
\begin{equation}
\hat{\underline{\theta}}_{\spacel,k} = \hat{\alpha}_{k} \hat{\underline{v}}_{\spacel,k'} + (\varU_{k})^2.
\end{equation} 

Recall that the background noise variance $(\varU_k)^2$ is known. Then, based on Eq.~\eqref{eq:scaleVarRelation}, an estimate of the broadband-plus-noise variance $\underline{\sigma}_{l,k}^2$ is given by
\begin{equation}\label{eq:varEstBroad}
\widehat{\underline{\sigma}^2_{\mathrlap{\spacel,k}}}\quad = \hat{\underline{\theta}}_{\spacel,k}.
\end{equation}
In what follows, it is assumed that the estimate $\widehat{\underline{\sigma}^2_{\mathrlap{\spacel,k}}}\hspace{2mm}$ for $k\in\Set{K}^{\symB}$ is perfect (no error), i.e., we use the variance estimate $\widehat{\underline{\sigma}^2_{\mathrlap{\spacel,k}}}$\hspace{2mm}  as the true variance $\underline{\sigma}^2_{\mathrlap{\spacel,k}}$\hspace{2mm}.

\subsection{Broadband and Tonal Components Parameter Estimation}\label{app:BTParameter}
\textbf{\textit{The Normalized Intensity Computation}}: At striation with index $l\in\SetStriation$ and frequency index $k \in \Set{K}^{\symBT}$, let $\underline{y}_{\spacel,k}^{\symBT} = \underline{x}^2_{\spacel,k}/(\underline{\sigma}_{\spacel,k}^2/2)$ be the normalized intensity. We compute the samples of the normalized intensity using the estimates in Eq.~\eqref{eq:varEstBroad} at the neighboring frequencies of $f_k\in\FBT$. By making use of the assumption that the standard deviation of the broadband component varies smoothly with frequency index $k$ as discussed in Sec.~\ref{sec:SigModel}, an estimate of $\underline{\sigma}^2_{\mathrlap{\spacel,k}}\hspace{2mm}$ for $k \in \Set{K}^{\text{bt}}$ is obtained\vspace{-1.5mm} as
\begin{equation}
\widehat{\underline{\sigma}^2_{\mathrlap{\spacel,k}}}\hspace{2mm} = \frac{1}{2} \Bigl(\hspace{.5mm}\widehat{\underline{\sigma}^2_{\mathrlap{\spacel,k-\Delta k}}}\qquad\hspace{1mm} + \widehat{\underline{\sigma}^2_{\mathrlap{\spacel,k+\Delta k}}}\qquad\hspace{1mm} \Bigr). \label{eq:VarianceEstimate}
\end{equation}
Here, the frequency index offset $\Delta k$ is chosen such that $k \pm \Delta k \in \Set{K}^{\symB}$, and thus, the estimates in Eq.~\eqref{eq:varEstBroad} can be utilized. In what follows, it is assumed that the estimate $\widehat{\underline{\sigma}^2_{\mathrlap{\spacel,k}}}\hspace{2mm}$ for $k\in\Set{K}^{\symBT}$ is perfect (no error). Thus, the computed normalized intensity values $\underline{y}^{\text{bt}}_{\spacel,k} = \underline{x}^2_{\spacel,k}/(\widehat{\underline{\sigma}^2_{\mathrlap{\spacel,k}}}\hspace{2mm}/2)\hspace{2mm}$ are samples of the PDF $\chi^2\big(\underline{y}^{\text{bt}}_{\spacel,k}; 2, \underline{\lambda}_{\spacel,k} \big)$ with the noncentrality parameter $\underline{\lambda}_{\spacel,k} = |\underline{p}_{\spacel,k}^{\symT}|^2/(\underline{\sigma}_{\spacel,k}^2)$ as discussed in Sec.~\ref{sec:SigModel}.

\textbf{\textit{Noncentrality Parameter Estimation}}: To compute the likelihood function using the samples of normalized intensity, we need to estimate the noncentrality parameters $\underline{\lambda}_{\spacel,k}$ of the normalized intensity. 
The numerator of the noncentrality parameter $|\underline{p}_{\spacel,k}^{\symT}|^2 = (m_{k}^{\symT}\hspace{.3mm}|\underline{g}_{\spacel,k}|)^2$ is the intensity of the tonal source signal modulated by the channel transfer function. 

Consistent with the variance estimation approach, under the transformation defined by the correct parameter vector $\V{q}$, the noncentrality parameter within a striation (across frequency indices $k\in\Set{K}^{\symBT}$), is determined by the frequency-dependent propagation effects of the channel and the magnitude of the source amplitude (cf.~definition of $|\underline{p}_{\spacel,k}^{\symT}|^2$ above as well as Eqs.~\eqref{eq:acousticME},~\eqref{eq:GScalarAssumption}, and~\eqref{eq:GScalarAssumptionTransformed}). While the noncentrality parameters $\underline{\lambda}_{\spacel,k}$ vary across striations, they are scaled versions of a common striation-dependent parameter at each frequency, meaning their relative change with frequency is consistent across all striations. This structure makes it possible to estimate the noncentrality parameters $\underline{\lambda}_{\spacel,k}$, based on normalized intensity measurements across both frequency and striation indices.

We estimate the noncentrality parameter $\underline{\lambda}_{\spacel,k}$\vspace{-2mm} as
\begin{equation}\label{eq:nc2Est}
\hat{\underline{\lambda}}_{\spacel,k} = \frac{\Bigl(\frac{1}{J}\Bigl(\sum_{i\in\Set{K}^{\symBT}} \underline{y}^{\symBT}_{\spacel,i}\Bigr)-2 \Bigr) \Bigl(\frac{1}{M} \Bigl( \sum_{j\in\SetStriation} \underline{y}^{\symBT}_{j,k}\Bigr)-2 \Bigr)}{\frac{1}{JM}\Bigl(\sum_{i\in\Set{K}^{\symBT}} \sum_{j\in\SetStriation} \underline{y}^{\symBT}_{i,j}\Bigr) - 2 }.
\vspace{0mm}
\end{equation}
A detailed derivation of this estimator is provided in the supplementary material\footnotemark[\value{footnote}].

\subsection{Range Estimation Algorithm}\label{app:estAlg}
The range estimation algorithm is outlined in Algorithm~\ref{algo:MLE_rA}.  We also introduce the normalized intensity matrix $\underline{\M{Y}}^{\symBT}$ of size $L\times J$ that is composed of the measurements $\underline{y}^{\symBT}_{\spacel,k}(\V{q})$. A preliminary step involves identifying and separating $\FB$ and $\FBT$. Given a frequency band of interest $\F$, the set of frequencies with tonal components ($\FBT$) consists of the peak frequencies of the power spectral density (PSD) estimates, and $\FB = \F\backslash\FBT$. The measurement matrix is then separated into $\underline{\M{X}}^{\symB}$ and $\underline{\M{X}}^{\symBT}$ accordingly. 

\setlength{\algomargin}{1pt}
\begin{algorithm}[t]
\caption{ML Range Estimation}\label{algo:MLE_rA}
\newcommand\mycommfont[1]{\small{#1}}               \SetCommentSty{mycommfont} \SetFuncSty{mycommfont}
\newcommand\myKWFont[1]{\textbf{\small{#1}}}        \SetKwSty{myKWFont}
\newcommand\myFuncArgFont[1]{\emph{\small{#1}}}     \SetFuncArgSty{myFuncArgFont} \SetProgSty{myFuncArgFont}
\newcommand\myArgStyFont[1]{\textsf{\small{#1}}}    \SetArgSty{myArgStyFont}

\SetKwInOut{Input}{Input}\SetKwInOut{Output}{Output}
\SetKwData{Beta}{$\hat{\beta}$}
\SetKwData{FreqB}{$\FB$}
\SetKwData{FreqBT}{$\FBT$}
\SetKwData{RadVel}{$\hat{\dot{\bm{r}}}$}
\SetKwData{FreqOffset}{$\Delta k$}
\SetKwData{VarBack}{$(\bm{\sigma}^{\symU})^2$}
\SetKwData{Sr}{$\Set{S}_{r}$}

\SetKwFunction{WhitenBroadbandSignal}{WhitenBroadbandSignal}
\SetKwFunction{NLTransform}{SampleIntensitiesAlongStriations}
\SetKwFunction{EstimateBVar}{EstimateBroadbandVariances}
\SetKwFunction{ComputeBroadbandScalars}{ComputeBroadbandScalars}
\SetKwFunction{ComputeISBR}{ComputeNormalizedIntensity}
\SetKwFunction{LoglikelihoodBroadband}{LoglikelihoodBroadband}
\SetKwFunction{Loglikelihood}{Loglikelihood}
\SetKwFunction{EstimateNCParam}{Estimate$\ist\chi^2\ist$Parameters}
\SetKwFunction{GetBroadband}{GetBroadbandComponents}
\SetKwFunction{GetBroadbandTonal}{GetBroadbandTonalComponents}

\KwData{A spectrogram $\M{X}$ of size $K \times N$}
\Input{\Sr, \Beta, \RadVel, \FreqB, \FreqBT, \VarBack}
\Output{$\rML$}
\For{$r \in \Sr$}{
    $\V{q} \leftarrow [r$, \Beta, \RadVel$]^{\text{T}}$\;
    $\underline{\M{X}}$ $\leftarrow$ \NLTransform($\M{X}$,\hspace{.6mm}$\V{q}$)\;
    $\underline{\M{X}}^{\symB}\leftarrow$\GetBroadband($\underline{\M{X}}$,\ist$\FreqB$)\;
    $\underline{\M{X}}^{\symBT}\leftarrow$\GetBroadbandTonal($\underline{\M{X}}$,\ist$\FreqBT$)\;

    \vspace{1mm}
    \tcc{Process measurements in $\FreqB$ (App.~\ref{app:broadParam})}
    $\underline{\hat{\V{\sigma}}}^2\leftarrow$ \EstimateBVar($\underline{\M{X}}^{\symB}$, \VarBack)\;

    \vspace{1mm}
    \tcc{Process measurements in $\FreqBT$ (App.~\ref{app:BTParameter})}
    $\underline{\M{Y}}^{\symBT}\leftarrow$\ComputeISBR($\underline{\M{X}}^{\symBT}$, $\underline{\hat{\V{\sigma}}}^2$)\;

    $\hat{\V{\underline{\lambda}}}\leftarrow$ \EstimateNCParam($\underline{\M{Y}}^{\symBT}$)\;
    
    \vspace{1mm}
    \tcc{Compute Log-likelihood}
    $LL(r) \leftarrow$ \Loglikelihood($\underline{\M{X}}^{\symB}$, $\underline{\M{Y}}^{\symBT}$, $\underline{\hat{\V{\sigma}}}^2$, $\hat{\V{\underline{\lambda}}}$)\; 
}
$\rML = \argmax_{r\in\Set{S}_{r}} LL(r)$;
\vspace{1mm}
\end{algorithm}

\section{Implementation Details} \label{app:impl}
\subsection{Spectrogram Computation}
Acoustic data were recorded at 25 kHz and downsampled to 2.5 kHz after low-pass filtering. Spectrograms were computed using 20-second Hamming windows with 50\% overlap, resulting in a frequency bin spacing of 0.05 Hz and snapshot spacing of 10 s. These settings were chosen to reduce spectral leakage from strong tonal components while maintaining sufficient temporal resolution. 

For each spectrogram segment for range estimation, a window of variable duration was selected to ensure at least $L=30$ valid striations (i.e., curves spanning the full processed frequency band) could be projected from candidate ranges. As the source moves farther from the receiver, the striations flatten, requiring longer temporal windows to maintain this coverage (see, for example, the pattern inside the panel N3 of Fig.~\ref{fig:spectrogramKD}).

\subsection{Ambient Noise}
Comparisons with ambient noise were performed using acoustic data from 11:00–11:30 (Mar 24\textsuperscript{th}), a period manually identified as having minimal ship activity based on AIS data. The maximum PSD during the quiet period in the 42–49 Hz band was 82 dB re: 1 \textmu Pa\textsuperscript{2}/Hz. As a reference, the minimum PSD during the event for MSC KALAMATA exceeded 90 dB re: 1 \textmu Pa\textsuperscript{2}/Hz. This also confirms that MSC KALAMATA's broadband emissions dominated over background noise. We use the PSD computed in this period as the additive background noise variance. 

\subsection{Frequency Band and Tonal Selection}
The processed frequency band varied across ship tracks to account for differences in radiated signal content and interference. For each spectrogram, prominent tonal frequencies were identified from power spectral density (PSD) peaks, forming the tonal set $\FBT$. Broadband bins $\FB$ were defined by selecting frequency bins at least 0.35 Hz away from any tonal component to minimize the potential influence of spectral leakage.

\subsection{Range Candidates}
For each processed spectrogram, candidate source ranges were defined on a uniform grid between $0.6 \cdot r_{\text{true}}$ and $1.4 \cdot r_{\text{true}}$ in 10 m increments, where $r_{\text{true}}$ is the AIS-derived ground truth range. Striations were projected using Eq.~\eqref{eq:wiStriation} with the region-specific WI value and assumed range rate. A fixed reference frequency is set as the center frequency of the considered frequency band for all projections.

\bibliography{IEEEabrv,StringDefinitions,SALBooks,SALPapers,Temp}
\end{document}